\documentclass[a4paper,11pt]{article}

\usepackage{jinstpub} 

\usepackage{lineno}
\usepackage{color}
\usepackage{bm}
\usepackage{here}
\usepackage{lscape}
\usepackage{comment}

\title{Characterization of Germanium Detectors for the Measurement of the Angular Distribution of Prompt $\gamma$-rays at the ANNRI in the MLF of the J-PARC}

\author[a]{Shusuke~Takada}
\author[b]{Takuya~Okudaira}
\author[b]{Fumiya~Goto}
\author[b]{Katsuya~Hirota}
\author[c]{Atsushi~Kimura}
\author[b]{Masaaki~Kitaguchi}
\author[a]{Jun~Koga}
\author[b]{Taro~Nakao}
\author[c]{Kenji~Sakai}
\author[b]{Hirohiko~M.~Shimizu}
\author[b]{Tomoki~Yamamoto}
\author[a]{Tamaki~Yoshioka}

\affiliation[a]{Kyushu University, 744 Motooka, Nishi, Fukuoka 819-0395, Japan}
\affiliation[b]{Nagoya University, Furocho, Chikusa, Nagoya 464-8602, Japan}
\affiliation[c]{Japan Atomic Energy Agency, 2-1 Shirane, Tokai 319-1195, Japan}

\abstract{In this study, the germanium detector assembly, installed at the Accurate Neutron-Nuclear Reaction measurement Instruments (ANNRI) in the Material and Life Science Facility (MLF) operated by the Japan Proton Accelerator Research Complex (J-PARC), has been characterized for extension to the measurement of the angular distribution of individual $\gamma$-ray transitions from neutron-induced compound states.
We have developed a Monte Carlo simulation code using the GEANT4 toolkit, which can reproduce the pulse-height spectra of $\gamma$-rays from radioactive sources and (n,$\gamma$) reactions. The simulation is applicable to the measurement of $\gamma$-rays in the energy region of 0.5--11.0 MeV.}

\keywords{Detector modelling and simulations I, Gamma detectors}

\begin{document}
\maketitle
\flushbottom

\section{Introduction}
A germanium detector assembly is installed at the Accurate Neutron-Nuclear Reaction measurement Instruments (ANNRI), located at the neutron beamline BL04 of the pulsed spallation neutron source of the Material and Life Science Facility (MLF) operated by the Japan Proton Accelerator Research Complex (J-PARC)~\cite{beamline}. 
The performance of the assembly enables the detection of prompt $\gamma$-rays \textcolor{black}{from (n,$\gamma$) reaction} with a sufficient energy resolution to resolve individual $\gamma$-ray transitions. 
The accuracy of the resonance parameters of the compound states has been improved by the combination of the $\gamma$-ray energy resolution and the capability to determine the incident neutron energy along with the neutron time-of-flight.

Extremely large parity violation (P-violation) was found in the helicity dependence of the neutron absorption cross section in the vicinity of the p-wave resonance for a variety of compound nuclei~\cite{lapvio}. 
The magnitude of the P-violating effects is $10^{6}$ times larger than that of the nucleon-nucleon interactions~\cite{pvio1,pvio2,pvio3}. 
The large P-violation is explained as an interference between the amplitudes of the p-wave resonance and the neighboring s-wave resonance~\cite{interference1,interference2}. 
The enhancement mechanism is expected to be applicable to P- and T-violating interactions and to enable highly sensitive studies of CP-violating interactions beyond the Standard Model of elementary particles~\cite{gudkov}. 
Theoretically, the interference between partial waves in the entrance channel causes an angular distribution of the individual $\gamma$-rays from the compound state~\cite{flambaum}. 
The ANNRI can measure the angular distribution of the $\gamma$-rays by using germanium detectors installed at a particular angle.

In this paper, we describe the determination of the pulse-height spectra of each germanium detector in the assembly by measuring the $\gamma$-rays from radioactive sources and $^{14}{\rm N}({\rm n},\gamma)$ reactions. 
We studied the $\gamma$-ray energy dependence of the detection efficiencies \textcolor{black}{and the angle of the germanium detectors} by simulations.

\section{Germanium Detector Assembly in the ANNRI}
\begin{figure}[htbp]
\centering\includegraphics[width=0.7\linewidth, angle=0]{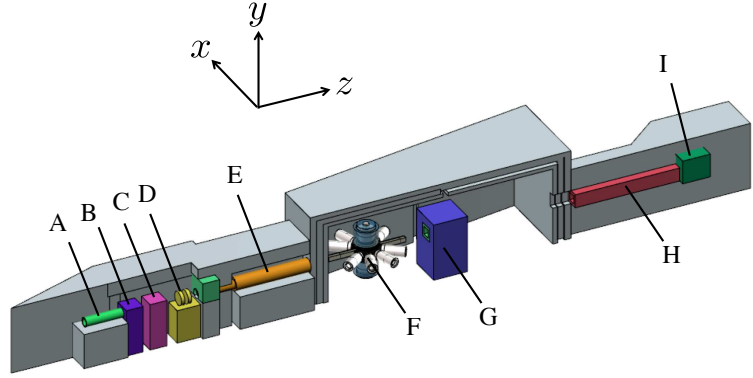}
\caption{\label{fig:beamline}
Schematic illustration of the ANNRI installed at the beamline BL04 of the MLF at the J-PARC (A) Collimator, (B) T0-chopper, (C) Neutron filter, (D) Disk chopper, (E) Collimator, (F) Germanium detector assembly, (G) Collimator, (H) Boron resin, and (I) Beam stopper (Iron).
}
\end{figure}

\begin{figure}[htbp]
\centering\includegraphics[width=0.7\linewidth, clip, angle=270]{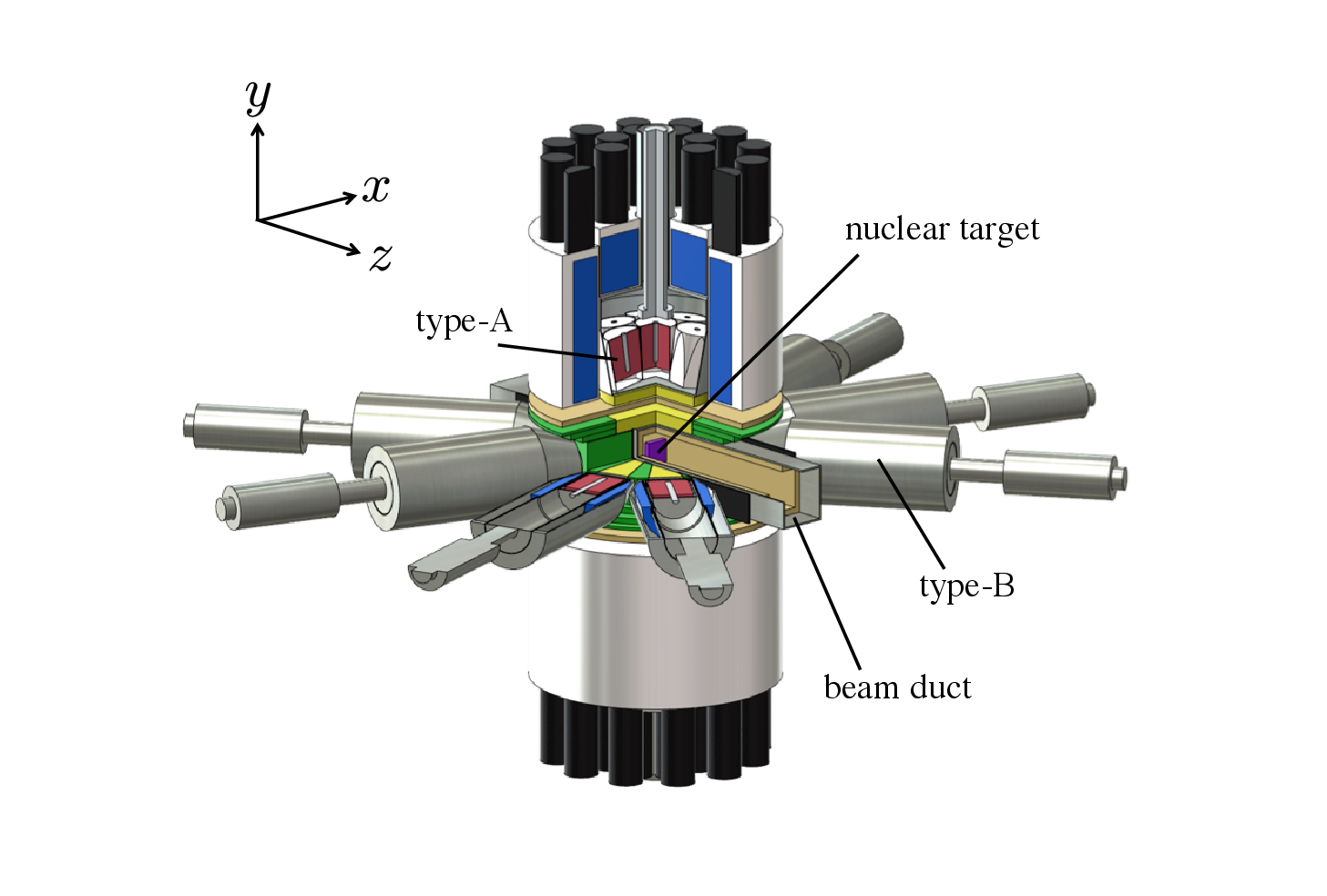}
\caption{\label{fig:detector}
Configuration of the germanium detector assembly.
}
\end{figure}

A schematic of the experimental setup of the ANNRI is shown in Fig.~\ref{fig:beamline}.
The configuration of the germanium detector assembly (F in Fig.~\ref{fig:beamline}) is shown in Fig.~\ref{fig:detector}.
The beam axis is represented by the $z$-axis, the vertical direction is the $y$-axis and the $x$-axis is defined such that they form a right-handed coordinate system.
The origin of the coordinate system is the position of the nuclear target.
The pulsed neutron beam is transported through a series of collimators (A in Fig.~\ref{fig:beamline}) to the target position located $21.5$ m from the moderator surface.
A T0-chopper (B in Fig.~\ref{fig:beamline}) and a disk chopper (D in Fig.~\ref{fig:beamline}) are installed at $12$ m and at $17$ m, to eliminate fast neutrons and cold neutrons, respectively~\cite{beamline}.
Several neutron filters (C in Fig.~\ref{fig:beamline}) are also inserted between the T0-chopper and the disk chopper, to adjust the beam intensity in the energy region of interest.

There are two kinds of germanium detectors: type-A and type-B.
The shape and dimension of type-A and type-B crystals are shown in Fig.~\ref{fig:Ge-A,Ge-B}.
\begin{figure}[htbp]
\centering\includegraphics[width=0.8\linewidth ]{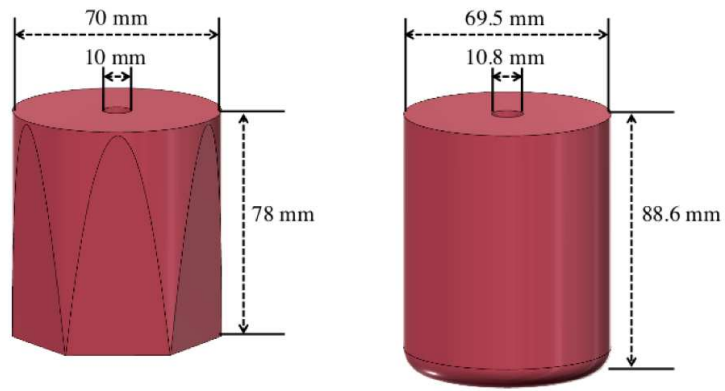}
\caption{\label{fig:Ge-A,Ge-B}
Schematics of type-A (left) and type-B (right) germanium crystals.
}
\end{figure}
The front-end of the type-A crystal has a hexagonal shape, and the back-end has a hole for the insertion of the electrode.
The seven type-A crystals form a detector unit as shown in Fig.~\ref{fig:cluster}.
Germanium detectors are covered by two neutron shields (F and H in Fig.~\ref{fig:cluster}) made of LiH with a thickness of 22.3 mm and 17.3 mm, respectively~\cite{detector}. The neutron shield G in Fig.~\ref{fig:cluster} is made of LiF and its thickness is 5 mm. 
The central crystal is directed toward the target center, while the surrounding six detectors are directed farther beyond the center. Therefore, they have different solid angles of $0.010 \times 4\pi \,{\rm sr}$ (central) and $0.0091 \times 4\pi \,{\rm sr}$ (one of the surrounding six detectors), respectively.
The side and back of the assembly of the type-A crystals are surrounded by bismuth germanate (BGO) scintillation detectors, illustrated by D in Fig.~\ref{fig:cluster}.
The polar angle in the spherical coordinate is denoted by $\theta$ and the azimuthal angle is $\varphi$.
Two assembly of the type-A crystals are placed at $(\theta,\varphi)=(90^\circ,90^\circ)$ and $(90^\circ,270^\circ)$. 
The central crystal of the upper (lower) type-A assembly is denoted by d1 (d8) and the other surrounding six detectors are denoted by d2--d7 (d9--d14), as shown in Table~\ref{tab:detector_angle}. 
\begin{figure}[htbp]
\centering\includegraphics[width=0.8\linewidth, clip, angle=0]{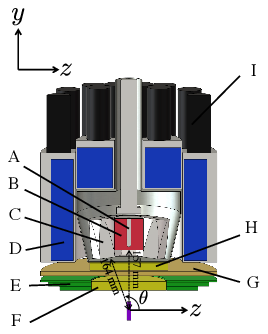}
\centering\includegraphics[width=0.8\linewidth, clip, angle=0]{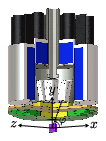}
\caption{\label{fig:cluster}
Schematic of a unit of type-A crystals consisting of seven type-A germanium detectors. Left figure: (A) Electrode, (B) Germanium crystal, (C) Aluminum case, (D) BGO crystal, (E) $\gamma$-ray shield (Pb collimator), (F) Neutron shield-1 (22.3 mm LiH), (G) Neutron shield-2 (5 mm LiF), (H) Neutron shield-3 (17.3 mm LiH), and (I) Photomultiplier tube for BGO crystal. Right figure shows the definition of the $\varphi$ angle.
}
\end{figure}

\begin{figure}[htbp]
\centering\includegraphics[width=0.7\linewidth, clip, angle=270]{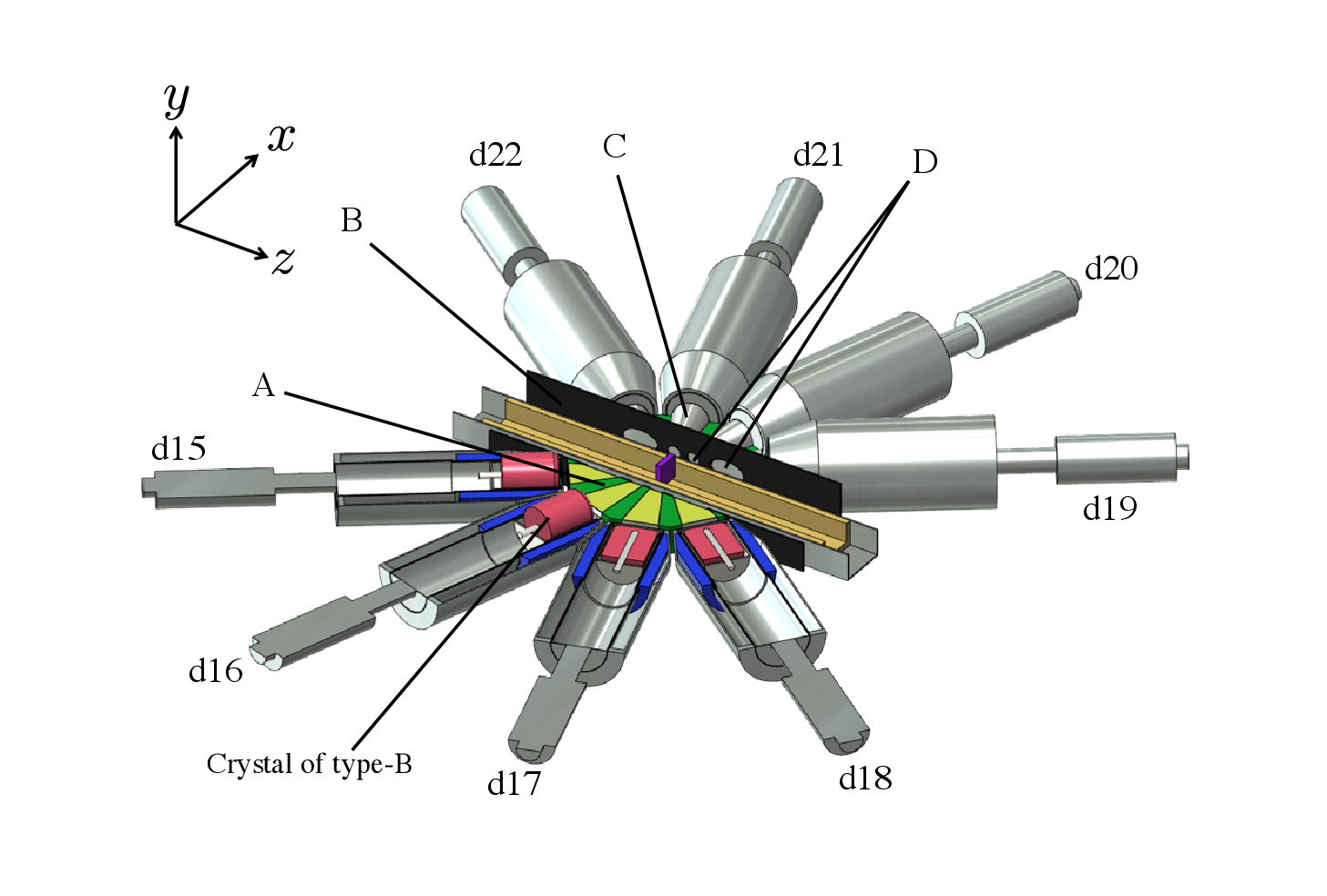}
\caption{\label{fig:coaxial8}
Schematic of the assembly of type-B crystals. (A) Pb collimator, (B) Carbon board, (C) LiH powder, and (D) the holes of collimator}
\end{figure}

\begin{figure}[htbp]
\centering\includegraphics[width=0.8\linewidth, clip, angle=0]{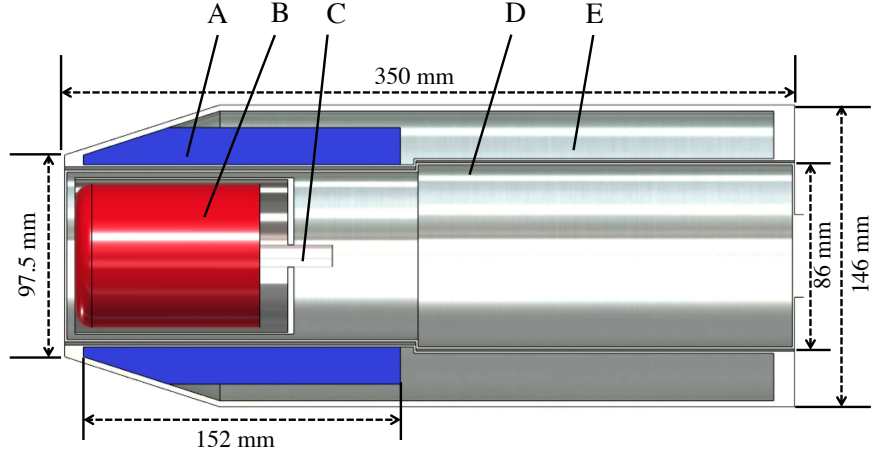}
\caption{\label{fig:coaxialunit}
Schematic of a unit of type-B crystals. (A) BGO crystal, (B) Germanium crystal, (C) Electrode, (D) and (E) Aluminum case. 
}
\end{figure}

The front-end of the type-B crystal has a circular shape and the back-end has a hole for the insertion of the electrode.
Eight type-B crystals are assembled, as shown in Fig.~\ref{fig:coaxial8}.
The surrounding of the type-B assembly is shown in Fig.\ref{fig:coaxialunit}.
Detectors are denoted by d15--d22.
It should be noted that the germanium crystal of detector d16 is smaller than that of the other detectors. 
All type-B crystals are directed toward the target center. 
Therefore, except for detector d16, they all have the same solid angle of $0.0072 \times 4\pi \,{\rm sr}$. 
The solid angle of detector d16 is $0.0048 \times 4\pi \,{\rm sr}$. 
Each type-B crystal is surrounded by a BGO scintillator (A in Fig.\ref{fig:coaxialunit}). 
A conical-shape $\gamma$-ray collimator made of Pb is located between each type-B crystal and the nuclear target. 
The diameter of the collimator at the front-end of the type-B crystal is 60 mm. 
The inside of the collimator is filled with LiF powder, which is encapsulated in an aluminum case, for absorbing the scattered neutrons.


The beam duct consists of two layers. 
The outer layer is made of aluminum with the thickness of 3 mm. 
The cross-sectional dimension is 86 mm $\times$ 96 mm. 
The inner layer is made of LiF with a thickness of 10.5 mm \textcolor{black}{for absorbing scattered neutrons}.

\section{Pulse-Height Spectra using Radioactive Sources and Melamine Target}
\begin{figure}[htbp]
\centering\includegraphics[width=0.7\linewidth]{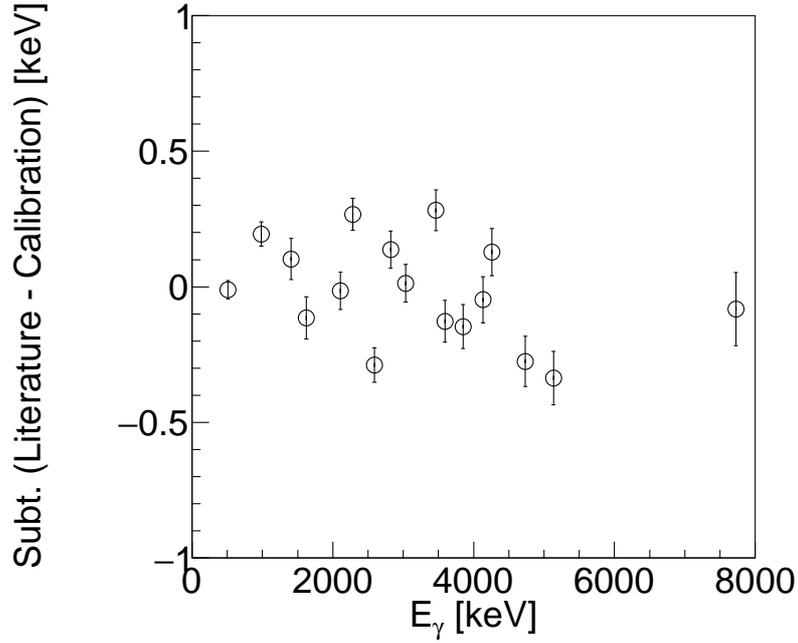}
\caption{\label{fig:linearity}
Verification of linearity for detector d10. The vertical axis is the ratio of $\gamma$-ray energy between the literature values~\cite{Al_gamma_energy} and calibrated values for the $^{27}$Al(n,$\gamma$) reaction. The horizontal axis is the energy of $\gamma$-ray.
}
\end{figure}

\textcolor{black}{The linearity of the detector response was ensured by measuring the pulse-height of prompt $\gamma$-rays from the neutron capture reactions with target nuclei of $^{27}$Al. 
We calibrated the $\gamma$-ray energy from the analog-to-digital converter (ADC) channel, by using 17 peaks of the $^{27}$Al(n,$\gamma$) reaction from 511 keV to 7724 keV.
The energy calibration was verified for several peaks of the $^{27}$Al(n,$\gamma$) reaction. 
The values on the vertical axis are expected to scatter around 0.
We confirmed good linearity with less than 1 keV of deviations, as shown in Fig.~\ref{fig:linearity}. 
The deviations are acceptable, as they are smaller than the energy resolution of the germanium detector.}

\subsection{Energy resolution of germanium detectors for $\gamma$-rays}
\begin{figure}[htbp]
\centering\includegraphics[width=0.6\linewidth]{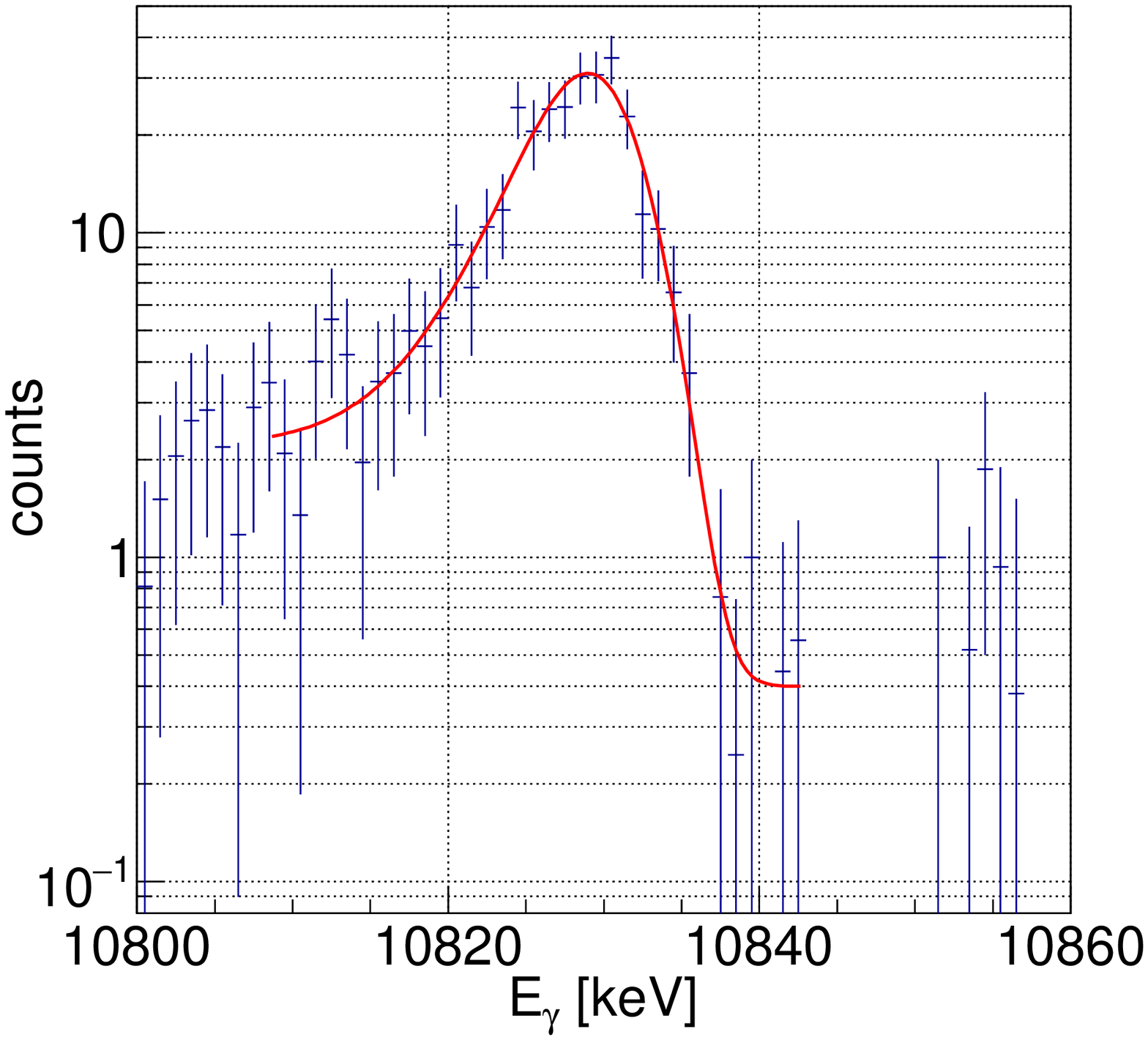}
\caption{\label{fig:gammaspec}
Shape of the full absorption peak of the 10829.11 keV $\gamma$-ray from the $^{14}$N(n,$\gamma$) reaction. The solid line represents the result of fitting with Eq.~\ref{eq:fitting}.
}
\end{figure}

\begin{figure}[htbp]
\centering\includegraphics[width=0.49\linewidth]{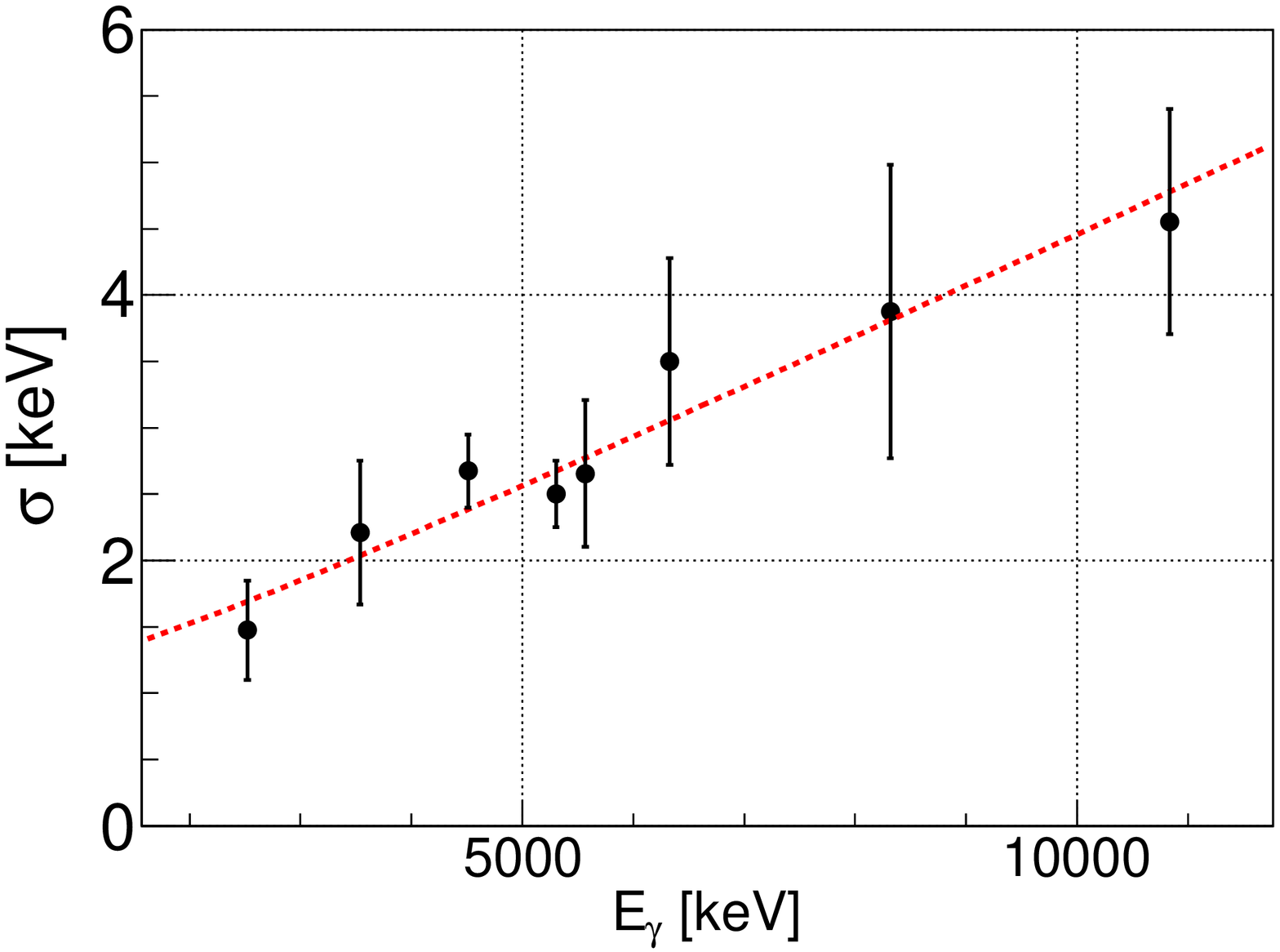}
\centering\includegraphics[width=0.49\linewidth]{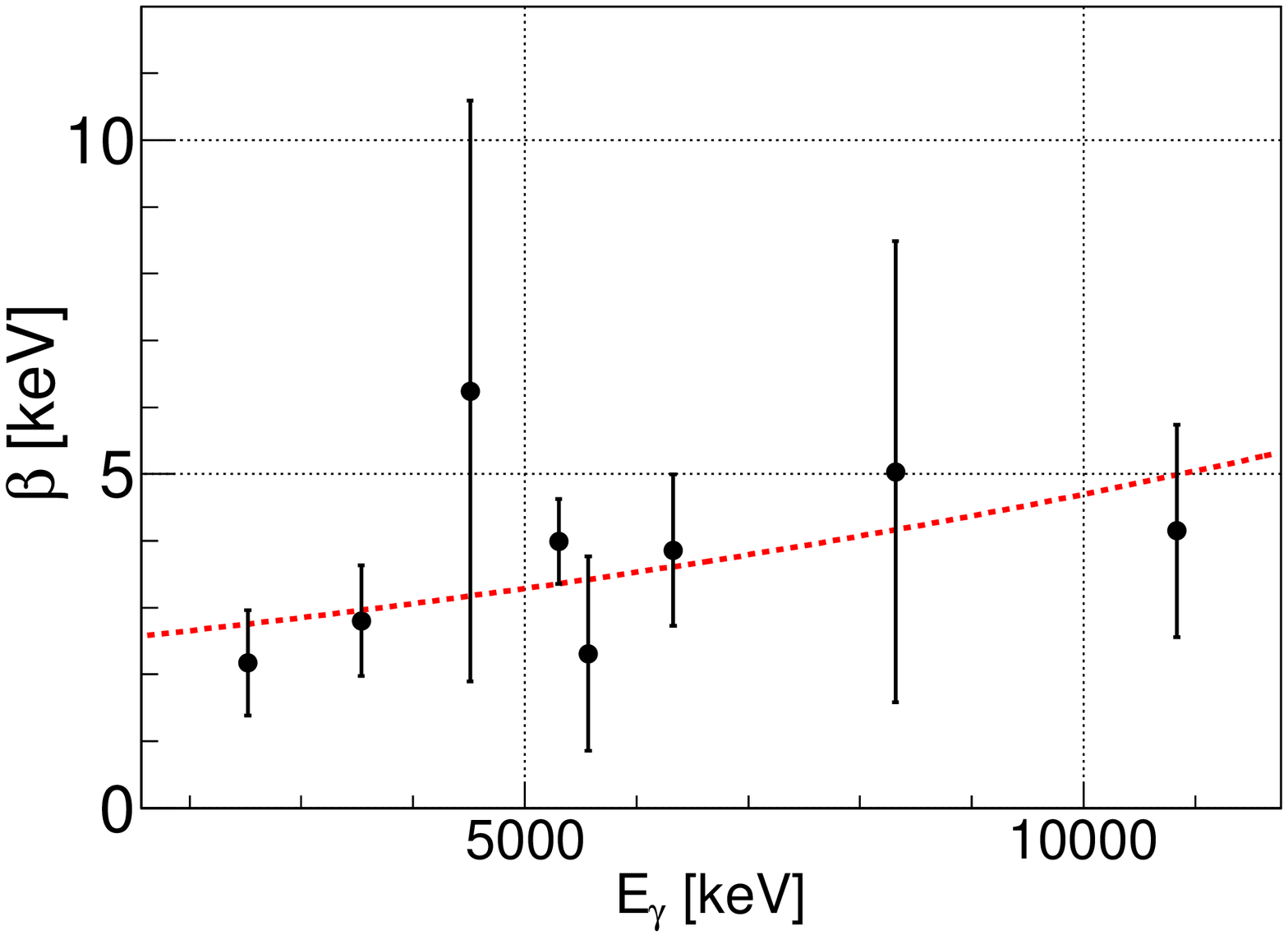} 
\caption{\label{fig:sigma_enedep}
Energy dependence of the obtained $\sigma$ (left) and $\beta$ (right) of detector d1.}
\end{figure}

We simulated the energy spectrum of the germanium detector assembly, by using GEANT4 9.6 ~\cite{geant}. 
The simulation of the energy resolution was implemented as follows. 
Ideally, the shape of the full absorption peak of the $\gamma$-ray is Gaussian, with a low energy tail (see Fig.~\ref{fig:gammaspec}).
Therefore the following function was used:
\begin{equation}
f(E) = F_{\rm gauss} + F_{\rm skew} + F_{\rm erfc},
\label{eq:fitting}
\end{equation}
where $F_{\rm gauss}, F_{\rm skew}$ and $F_{\rm erfc}$ are the Gaussian function, skewed Gaussian function, and the complementary error function, respectively. These are expressed as follows:
\begin{eqnarray}
F_{\rm gauss} &=& {C} \exp{\left(- \left(\frac{E-c}{\sqrt{2}\sigma} \right)^2 \right)}, \\
F_{\rm skew} &=& {D}\exp{\left( \frac{E - c}{\beta} \right)}\ {\rm erfc}\left( \frac{E-c}{\sqrt{2}\sigma} - \frac{\sigma}{\sqrt{2}\beta} \right), \\
F_{\rm erfc} &=& {G}~{\rm erfc}\left( \frac{E-c}{\sqrt{2}\sigma} \right) + {H},
\end{eqnarray}
where $c$ is the peak position, $\sigma$ is the standard deviation of $F_{\rm gauss}$, and $\beta$ represents the extent of the low energy tail. 
Functions $F_{\rm skew}$ and $F_{\rm erfc}$ are introduced to account for the low energy tail.
The obtained $\sigma$ and $\beta$ as a function of the $\gamma$-ray energy are shown in Fig.~\ref{fig:sigma_enedep}.
The obtained $\sigma$ as a function of the energy is fitted by using the following formula:
\begin{equation}
\sigma(E) = \sqrt{\sigma_D^2 + \sigma_X^2 + \sigma_E^2},
\end{equation}
where $\sigma_D, \sigma_X$, and $\sigma_E$ denote the Fano statistics, collection efficiency of charge carriers, and electrical noise, respectively~\cite{resolution}. These are defined as follows:
\begin{eqnarray}
\sigma_D &=& \sqrt{F\epsilon E}, \\
\sigma_X &=& {A} E, \\
\sigma_E &=& {B},
\end{eqnarray}
where $F$ is the Fano factor ($F =0.112 \pm 0.001$~\cite{fano}) and $\epsilon$ is the energy of the creation of an electron-hole pair in the germanium. 
The obtained $\beta$ as a function of the energy is fitted by the following empirical formula:
\begin{equation}
\beta(E) = {K}\exp{({L} E)}.
\end{equation}
We obtained $A=(3.94\pm0.68) \times 10^{4}$, $B=1.02 \pm 0.62$, $K=2.30\pm0.61$, and $L=(7.13 \pm 4.37) \times 10^{-5}$ as results from the fitting, as shown in Fig.~\ref{fig:sigma_enedep}.
The parameters for $\sigma$ and $\beta$ were implemented in the simulation.

\subsection{Comparisons of measurement and Monte Carlo simulation in the order of 1 MeV}
\begin{figure}[htbp]
\centering\includegraphics[width=0.47\linewidth]{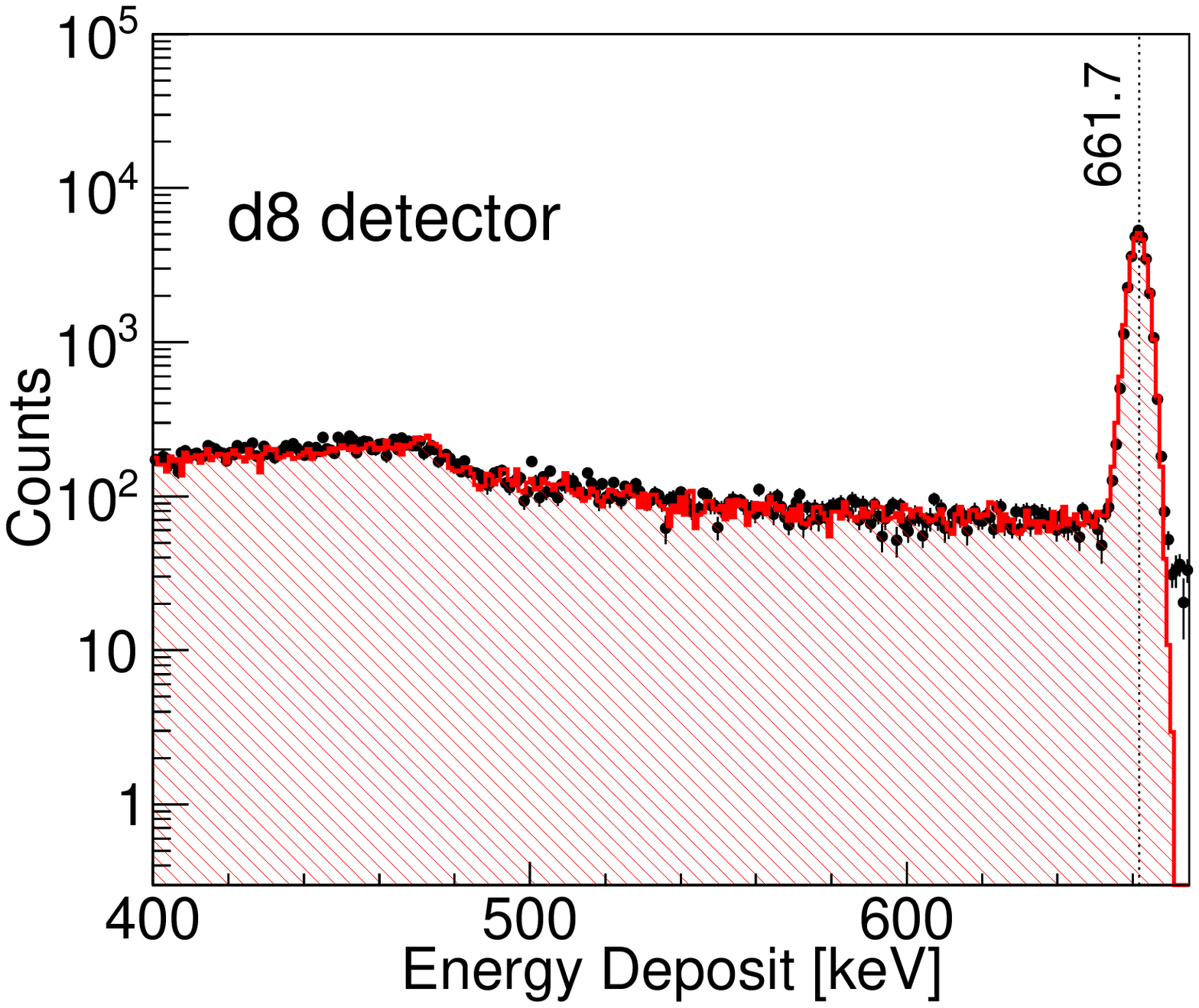}
\centering\includegraphics[width=0.47\linewidth]{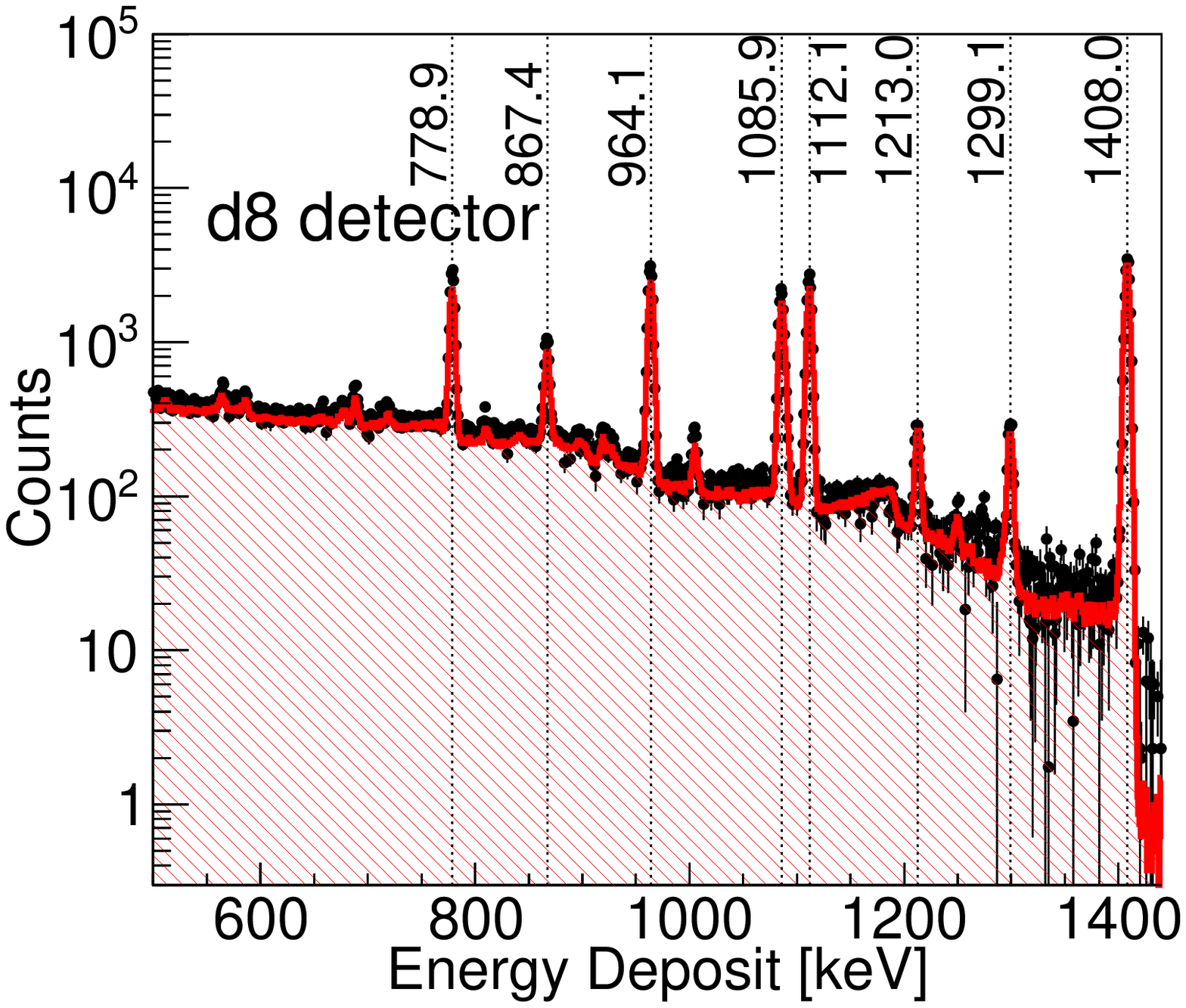}
\centering\includegraphics[width=0.47\linewidth]{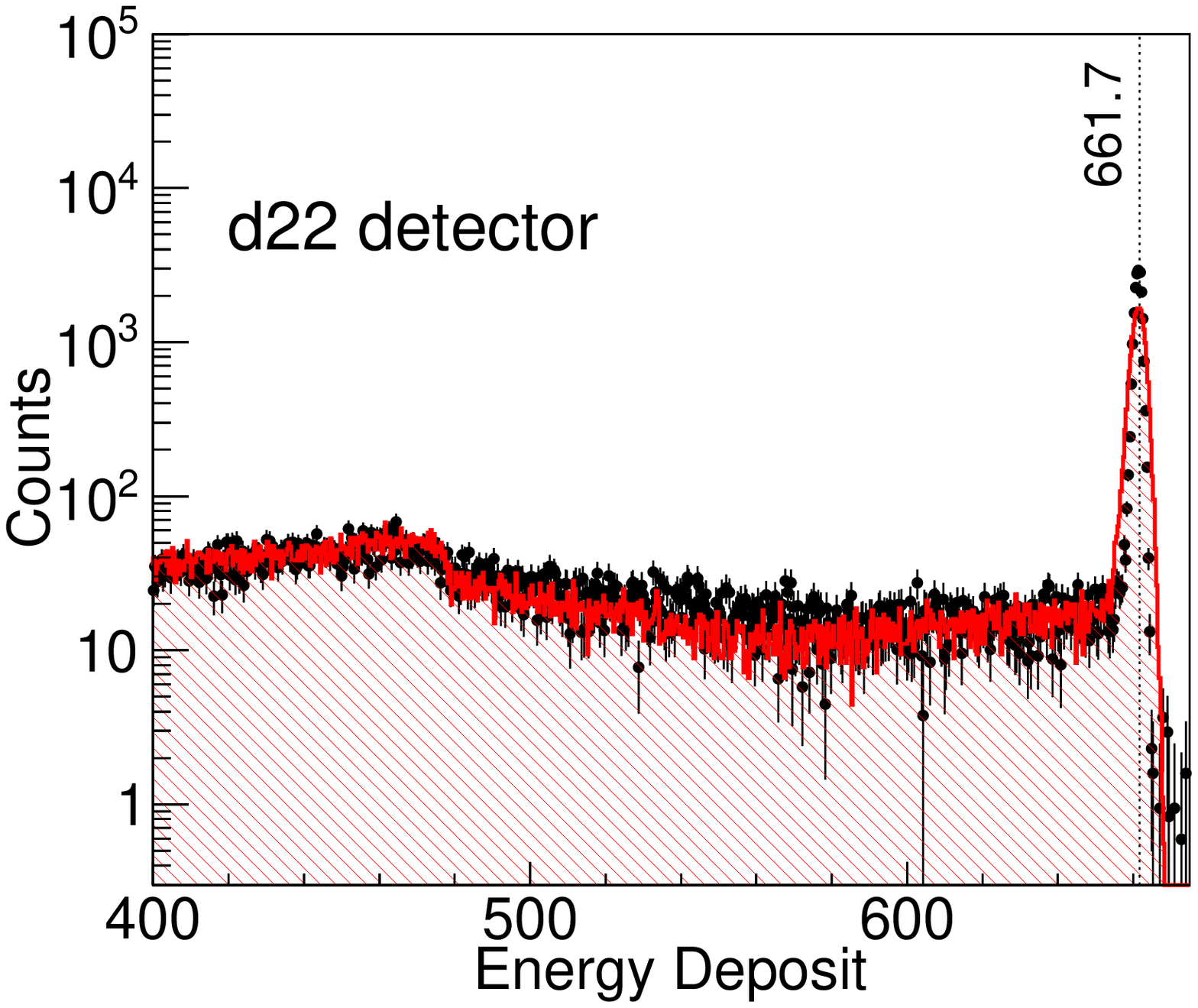}
\centering\includegraphics[width=0.47\linewidth]{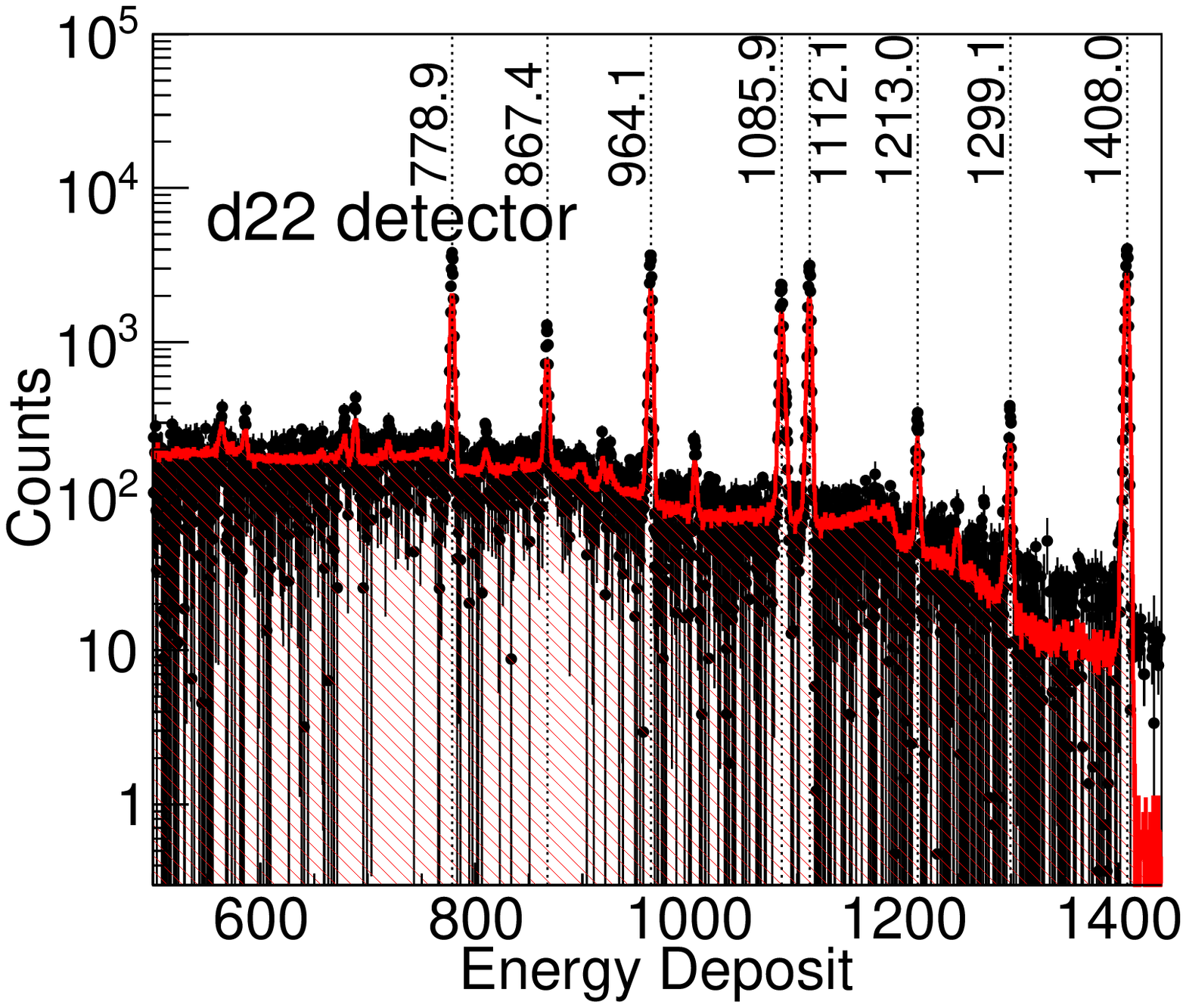}
\centering\includegraphics[width=0.47\linewidth]{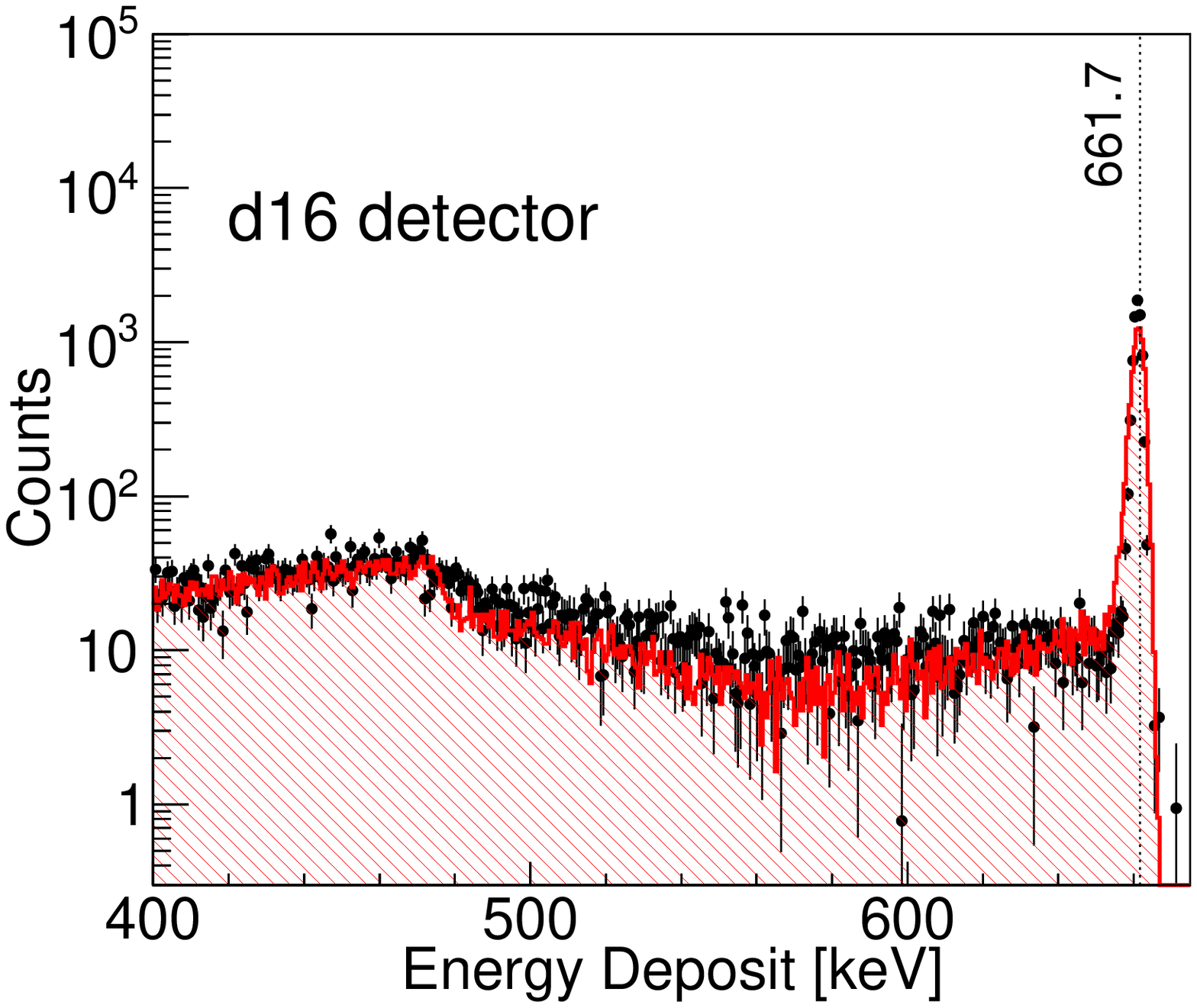}
\centering\includegraphics[width=0.47\linewidth]{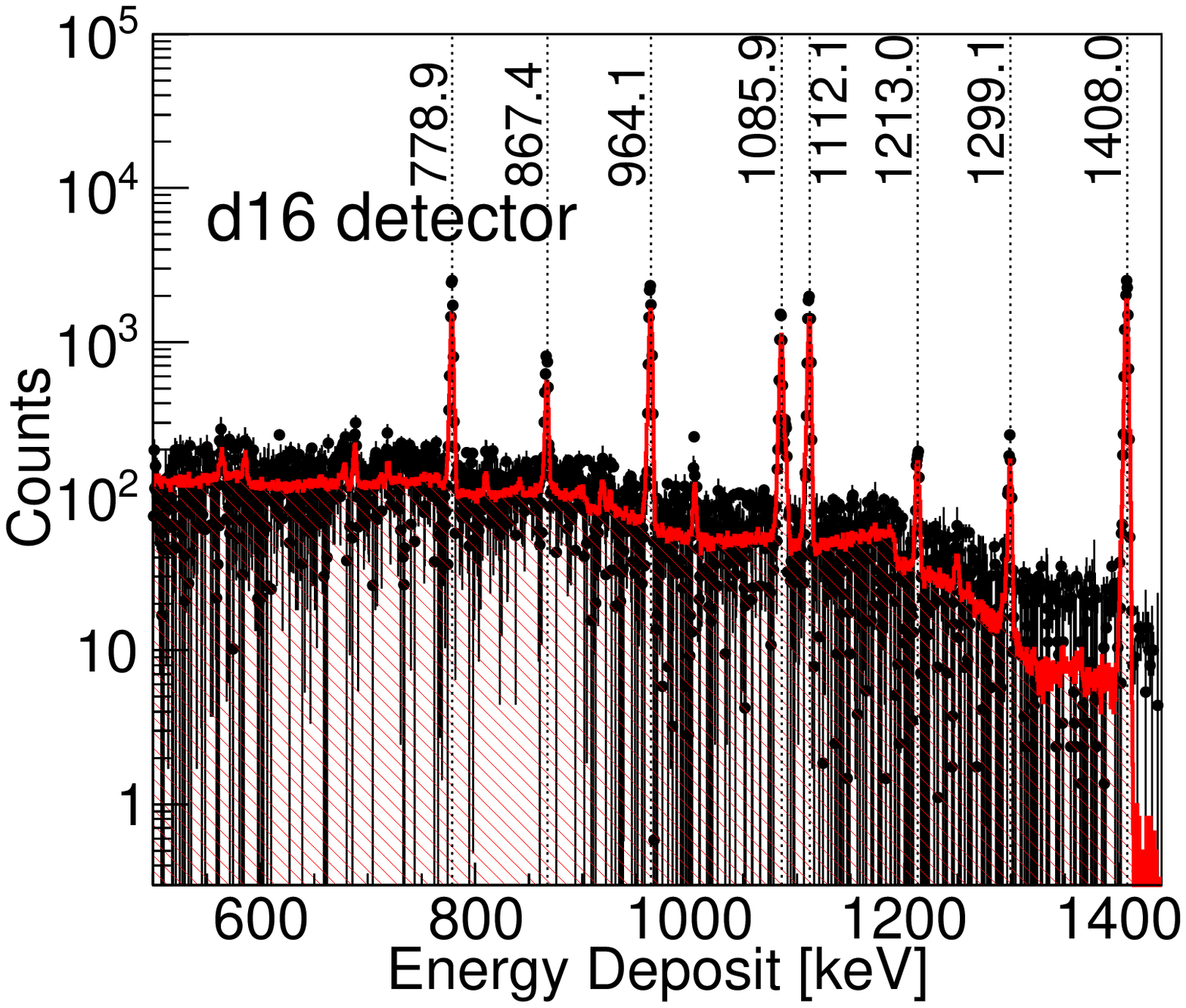}
\caption{\label{fig:source-spectra}
Pulse-height spectra of $\gamma$-rays from $^{137}$Cs (left) and $^{152}$Eu (right) radioactive sources. 
The spectra were measured by detectors d8, d22, and d16, shown in the top, middle, and bottom figures, respectively. 
The black dots and shaded histogram represent the measurement and the simulation, respectively. 
}
\end{figure}

In order to confirm the reproducibility of the simulation, we measured the energy spectra of $\gamma$-rays of the germanium detector from $^{137}$Cs and $^{152}$Eu radioactive sources placed at the nuclear target position.
The spectra obtained are shown in Fig.~\ref{fig:source-spectra} represented by black dots, together with the simulated spectra.
Each simulated pulse-height spectrum was adjusted to fit to the corresponding measured spectrum with a single parameter of the average efficiency, as shown in Fig.~\ref{fig:source-spectra}.

\subsection{Comparison of measurement and Monte Carlo simulation up to 11 MeV}
The prompt $\gamma$-rays from the (n,$\gamma$) reaction have $\gamma$-ray energies of approximately 5--9 MeV. Therefore, the $\gamma$-ray from the ${^{14}{\rm N}({\rm n},\gamma)}$ reaction with a $\gamma$-ray energy of up to 11 MeV is useful to verify the reproducibility of the simulation. 
The $\gamma$-ray energy spectrum of the ${^{14}{\rm N}({\rm n},\gamma)}$ reaction was obtained by using a melamine target with a diameter of 10.75 mm and a thickness of 1 mm. 

\begin{figure}[htbp]
\centering\includegraphics[width=0.8\linewidth]{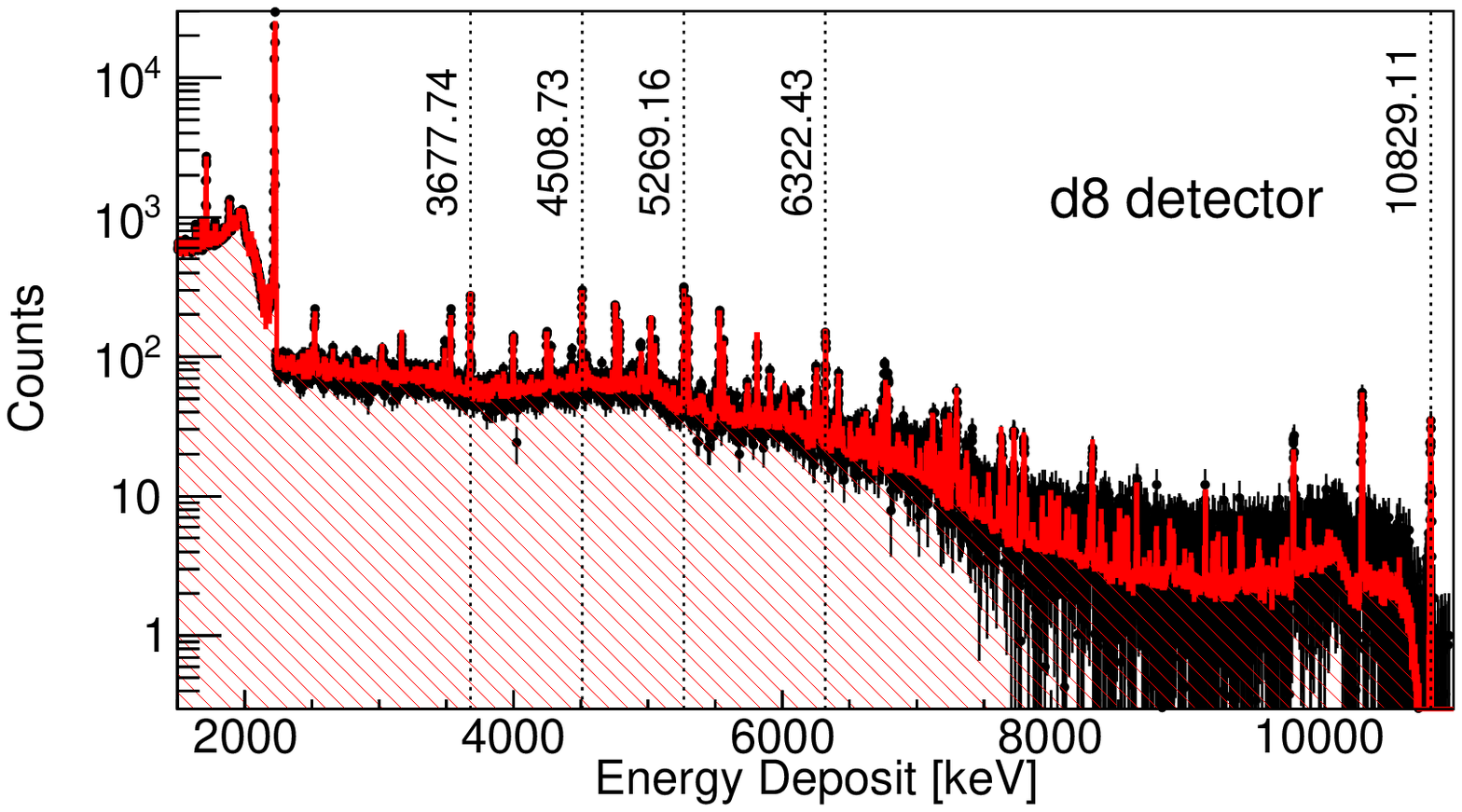}
\centering\includegraphics[width=0.8\linewidth]{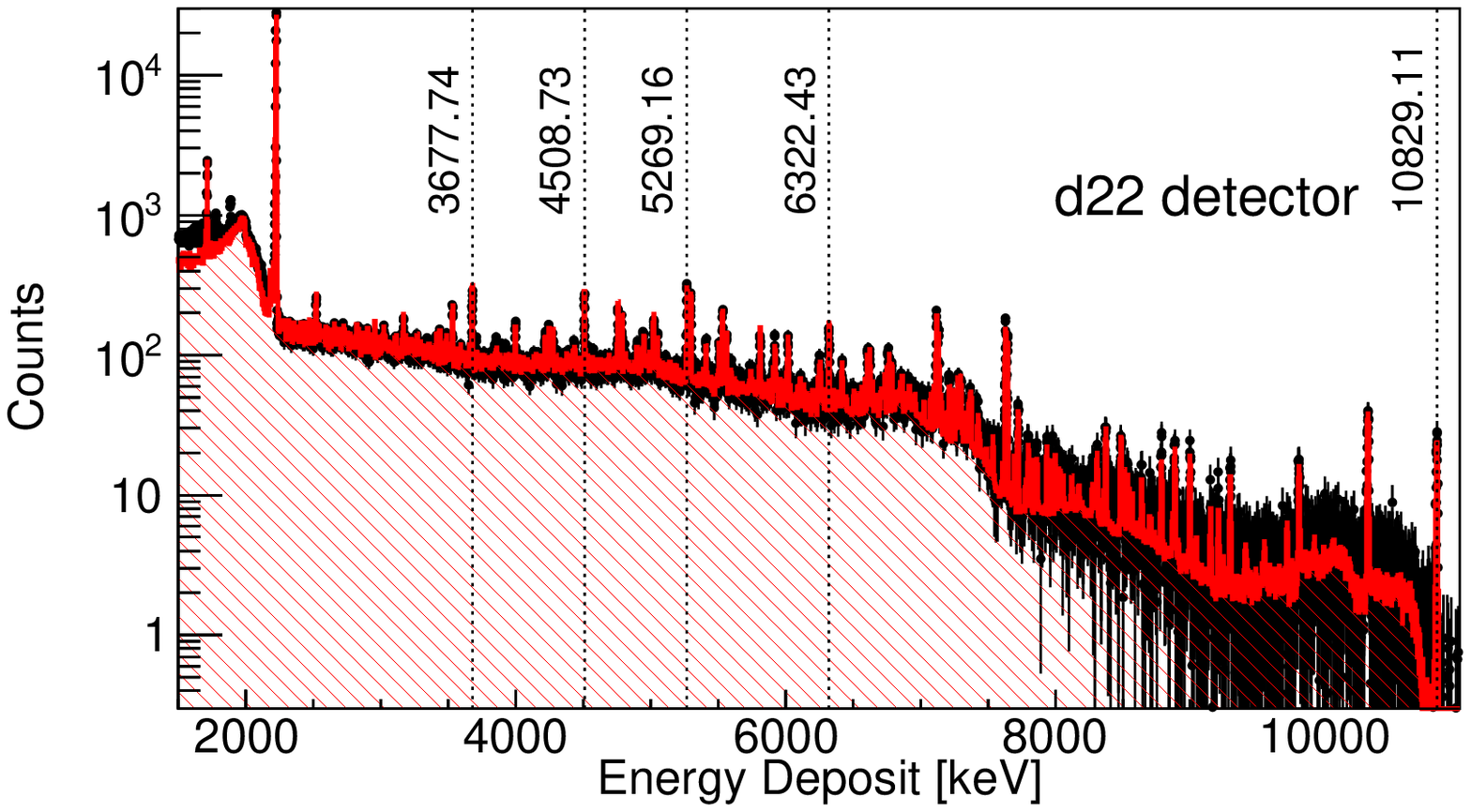}
\centering\includegraphics[width=0.8\linewidth]{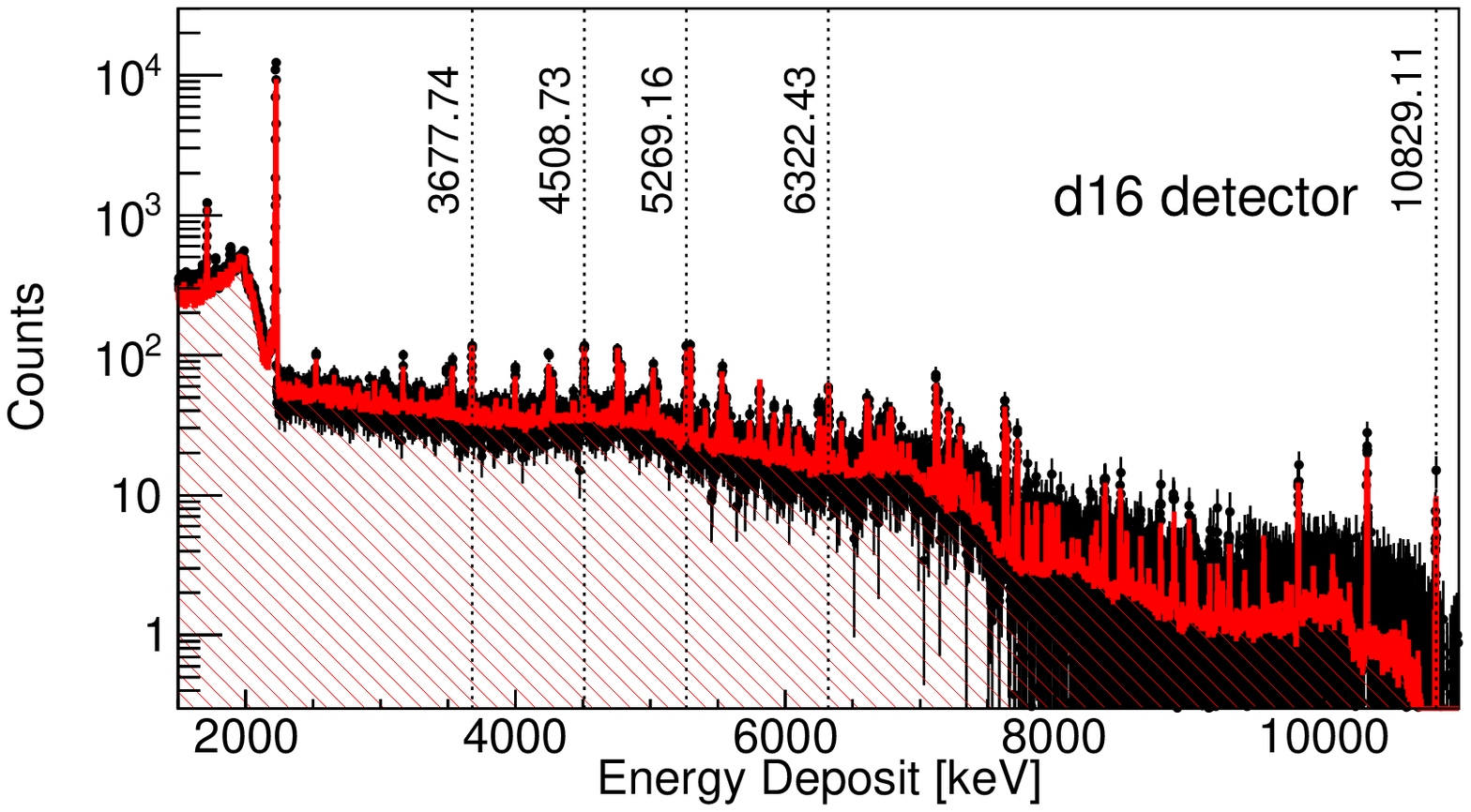}
\caption{\label{fig:melamine-spectrum}
Pulse-height spectra of $\gamma$-rays from the $({\rm n},\gamma)$ reaction of the melamine target. 
The spectra were measured by detectors d8, d22, and d16 are shown in the top, middle, and bottom figures, respectively. 
The black dots and shaded histogram represent the data and simulation, respectively. 
}
\end{figure}

Figure \ref{fig:melamine-spectrum} shows comparisons of the melamine measurements and simulations in the detectors d8, d22, and d16.
The spectrum of the measurement was reproduced by summing up the simulated spectra of monochromatic $\gamma$-rays.
Here, in addition to $\gamma$-rays from C, H, and N with melamine target, Li, F, Al, Fe, and Ni were also considered.

\begin{table}[htbp]
	\centering
	\begin{tabular}{|c || c | c | c | c |}
	\hline
	detector ID & $\theta$ [deg] & $\varphi$ [deg] \\
	\hline
	d1	&	$90.0$	&	$90.0$	\\
	d2	&	$90.0$	&	$66.3$	\\
	d3	&	$70.9$	&	$78.2$	\\
	d4	&	$70.9$	&	$101.8$	\\
	d5	&	$90.0$	&	$113.7$	\\
	d6	&	$109.1$	&	$101.8$	\\
	d7	&	$109.1$	&	$78.2$	\\
	d8	&	$90.0$	&	$270.0$	\\
	d9	&	$90.0$	&	$293.7$	\\
	d10	&	$70.9$	&	$281.8$	\\
	d11	&	$70.9$	&	$258.2$	\\
	d12	&	$90.0$	&	$246.3$	\\
	d13	&	$109.1$	&	$258.2$	\\
	d14	&	$109.1$	&	$281.8$	\\
	d15	&	$144.0$	&	$180.0$	\\
	d16	&	$108.0$	&	$180.0$	\\
	d17	&	$72.0$	&	$180.0$	\\
	d18	&	$36.0$	&	$180.0$	\\
	d19	&	$36.0$	&	$0.0$	\\
	d20	&	$72.0$	&	$0.0$	\\
	d21	&	$108.0$	&	$0.0$	\\
	d22	&	$144.0$	&	$0.0$	\\
	\hline
	\end{tabular}
	\caption{
	 $\theta$ and $\varphi$ are the angles at the center of the front-end of each detector.}
	\label{tab:detector_angle} 	
\end{table}

\section{Peak Efficiency of Individual Germanium Detectors}
\begin{figure}[htbp]
\centering\includegraphics[width=0.8\linewidth, angle=0]{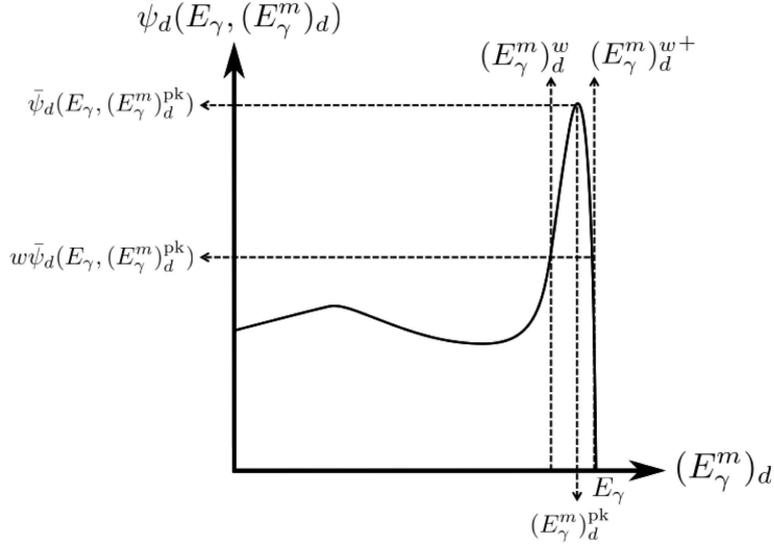}
\caption{
Definition of the full absorption peak in the pulse-height spectrum.
}
\label{fig:defspec}
\end{figure}

We define the probability of a $\gamma$-ray emitted to the direction of $\Omega_\gamma=(\theta_\gamma,\varphi_\gamma)$ from the target position with a $\gamma$-ray energy of $E=E_\gamma$ at the $d$th detector as $\psi_d(E_\gamma,\Omega_\gamma,(E^{{\rm m}}_{\gamma})_d)$.
The distribution of the energy deposit is defined as
\begin{equation}
\bar{\psi_d}(E_\gamma,(E^{{\rm m}}_{\gamma})_d) = \int^{}_{\Omega_d} \psi_d(E_\gamma,\Omega_\gamma,(E^{{\rm m}}_{\gamma})_d) \ d\Omega_\gamma,
\end{equation}
where $\Omega_d$ is the geometric solid angle of the $d$th detector.
The efficiency of the emitted $\gamma$-rays at each detector is defined as
\begin{equation}
\epsilon^{{\rm pk},w}_{d}(E_\gamma)= \int^{{(E^{{\rm m}}_{\gamma})^{w}_{d}}^{+}}_{{(E^{{\rm m}}_{\gamma})^{w}_{d}}^{-}} \bar{\psi_d}(E_\gamma,(E^{m}_{\gamma})_d) \ d(E^{{\rm m}}_\gamma)_d,
\label{eq:epsilon_peak}	
\end{equation}
where ${(E^{{\rm m}}_{\gamma})^{w}_{d}}^{+}$ and ${(E^{{\rm m}}_{\gamma})^{w}_{d}}^{-}$ are the upper and lower limits for the region of the full absorption peak. We used $w=1/10$ (full width of 10th maximum of the peak of the probability) as an integration interval. These definitions are summarized in Fig.~\ref{fig:defspec}.

The $\gamma$-ray distribution can be expanded by a sum of Legendre polynominals $\sum^{\infty}_{p=0} c_p P_p({\rm cos} \ \theta_\gamma)$ as
\begin{eqnarray}
N_d(E_{\gamma})=N_0 \sum^{\infty}_{p=0} c_p \bar{P}_{d,p},
\end{eqnarray}
\begin{equation}
\bar{P}_{d,p}=\frac{1}{4 \pi} \int^{{(E^{{\rm m}}_{\gamma})^{w}_{d}}^{+}}_{{(E^{{\rm m}}_{\gamma})^{w}_{d}}^{-}} {\rm d}(E^{\rm m}_{\gamma})_d \ \int d\Omega_{\gamma} P_p({\rm cos} \ \theta_{\gamma})\bar{\psi}_d(E\gamma,\Omega_\gamma),
\end{equation}
\begin{equation}
\psi^{\prime}_d(\cos\theta,E_{\gamma})=\int^{}_{\varphi_\gamma} \psi_d(E_\gamma,\Omega_\gamma,(E^{{\rm m}}_{\gamma})_d) \ d\varphi_\gamma.
\end{equation}

\begin{figure}[htbp]
\centering\includegraphics[width=0.49\linewidth, angle=0]{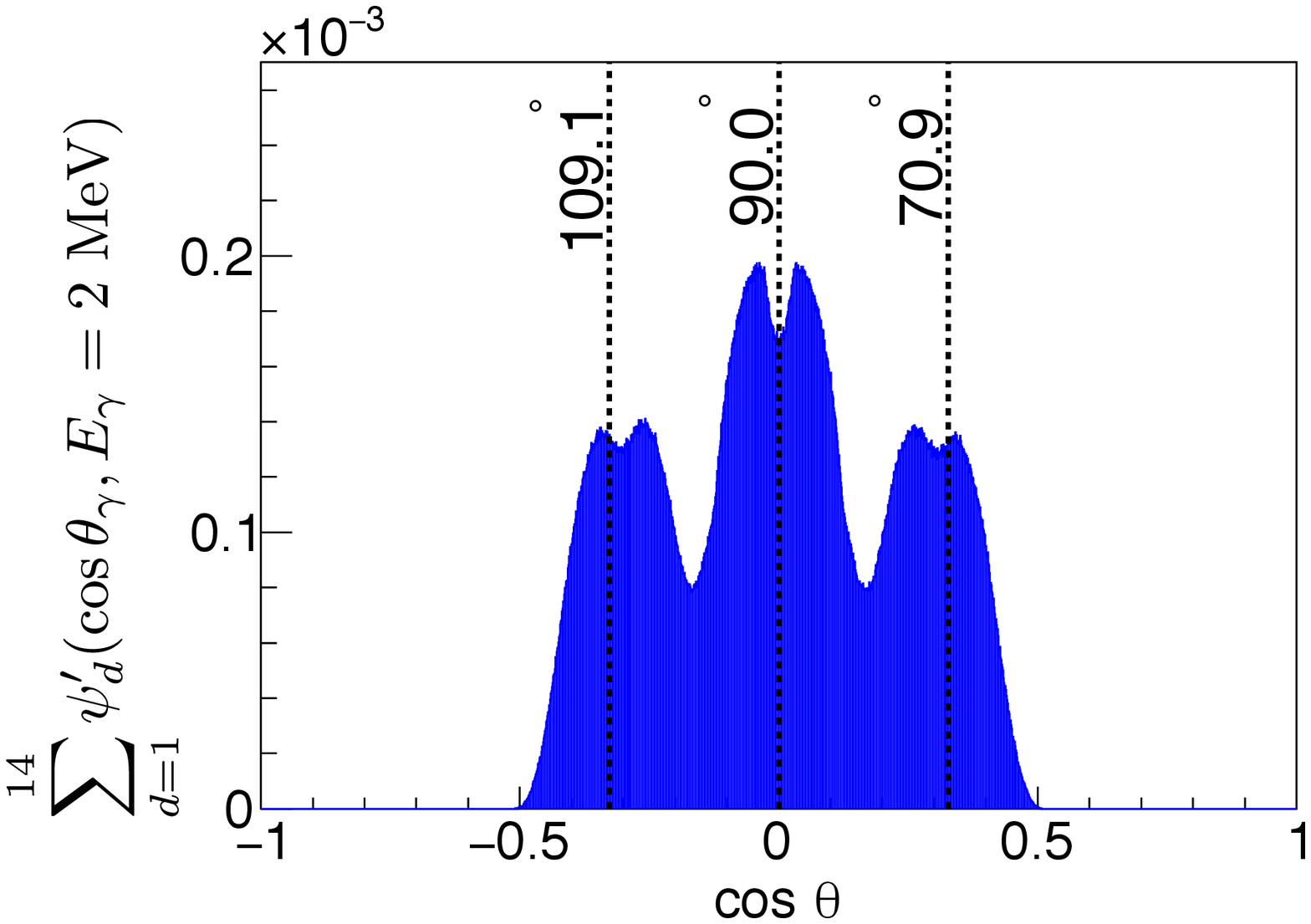}
\centering\includegraphics[width=0.49\linewidth, angle=0]{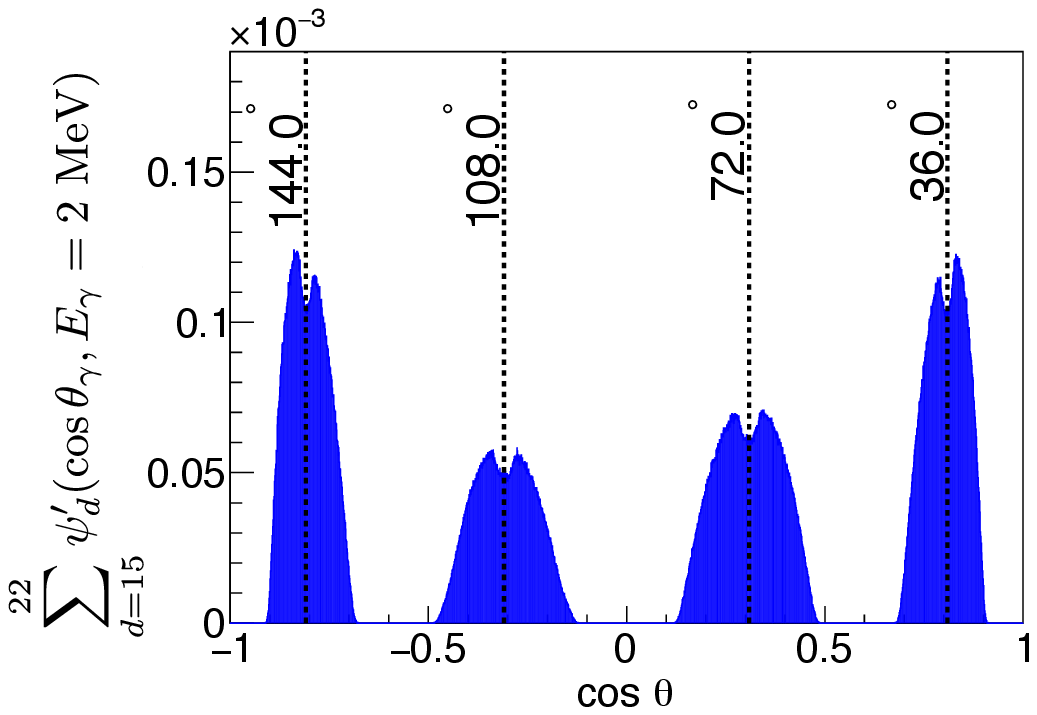}
\caption{\label{fig:psi}
Values of $\psi_i(\cos\theta,E_\gamma)$ as a function of $\cos\theta$ for $E=2$ MeV for the type-A (left) and type-B (right) assemblies.
}
\end{figure}

The function $\psi^{\prime}_d(\cos\theta,E_\gamma)$ was calculated by the simulation, as shown in Fig.~\ref{fig:psi}.
The dip structure of the peak shown in Fig.~\ref{fig:psi} is due to the opening on the back of germanium crystal where the electrode is inserted.
The vertical dotted lines in Fig.~\ref{fig:psi} represent the angle of the detector center as indicated in Table~{\ref{tab:detector_angle}}.

\begin{figure}[htbp]
\centering\includegraphics[width=0.6\linewidth, angle=270]{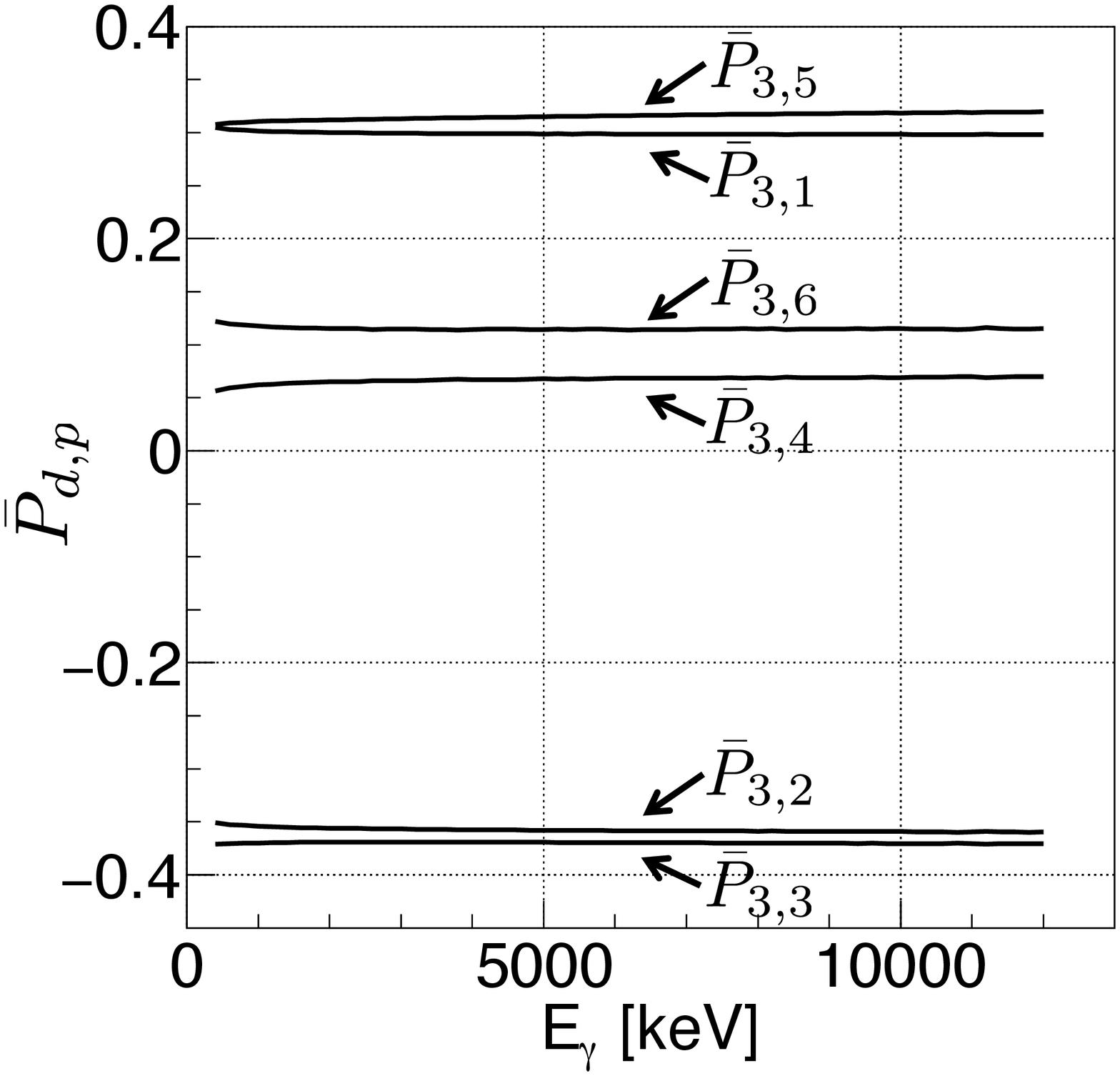}
\caption{\label{fig:pil}
Value of $\bar{P}_{d,p}$ as a function of $E_\gamma$ for detector d3.
}
\end{figure}

In the left figure in Fig.~\ref{fig:psi}, the dotted lines for ${\theta=70.9^{\circ}}$ and ${\theta=109.1^{\circ}}$ deviate from the center of each front-face, as the surrounding six detectors in the type-A assembly are not directed toward the target.
This effect is more visible as $E_\gamma$ becomes higher.
In the right figure in Fig.~\ref{fig:psi}, the peak shapes for ${\theta=36.0^{\circ}}$ and ${\theta=144.0^{\circ}}$ are different from the others due to the different acceptance angles of the collimator (see D in Fig.~\ref{fig:coaxial8}).
The peak for ${\theta=108.0^{\circ}}$ is smaller than that of ${\theta=72.0^{\circ}}$, due to the size of detector d16.
The calculated $\epsilon^{{\rm pk},w}_{d}(E_\gamma)$ and $\bar{P}_{d,p}(E_\gamma)$ values are listed in Table~\ref{jikkou1MeV}-\ref{jikkou8MeV} for $1 \le p \le 6$ and shown in Fig.~\ref{fig:pil} for detector d3.

\begin{figure}[htbp]
\centering\includegraphics[width=0.6\linewidth, angle=270]{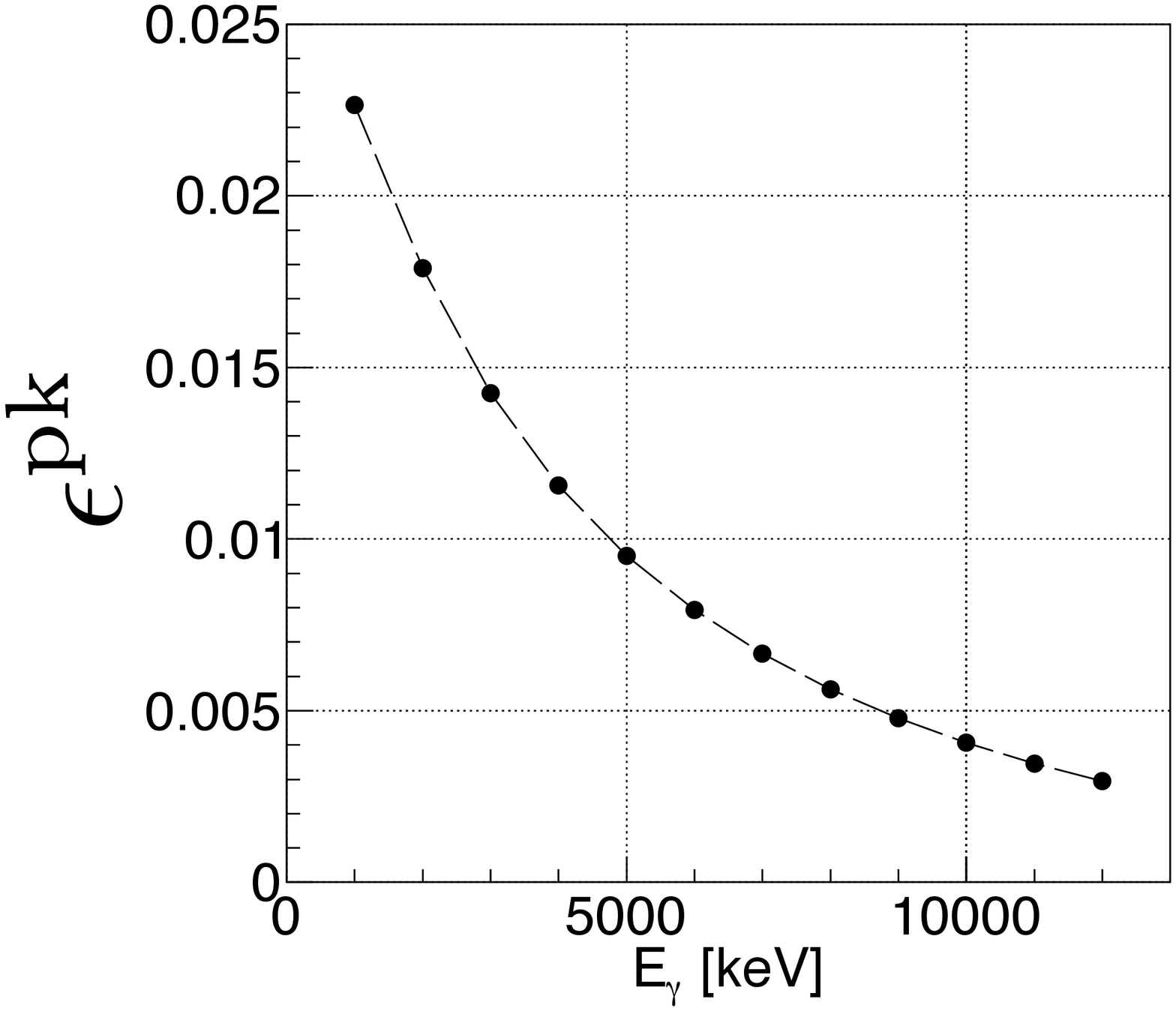}
\caption{\label{fig:epsilon_ene}
$\epsilon^{{\rm pk}} $ as a function of $E_\gamma$.
}
\end{figure}

In the $\epsilon^{{\rm pk},w}_{d}(E_\gamma)$ column, the values are decreasing as the $\gamma$-ray energy increases, due to the punch-through effect.
Nevertheless, the deviations from the averaged $\epsilon^{{\rm pk},w}_{d}(E_\gamma)$ values also decrease as the $\gamma$-ray energy increases, because the Compton scattering due to the materials between the target and the germanium detector is smaller for higher-energy $\gamma$-rays.
Figure \ref{fig:epsilon_ene} shows the energy dependence of $\epsilon^{\rm pk} $. 
Here, $\epsilon^{\rm pk} $ is defined as follows
\begin{equation}
\epsilon^{\rm pk} = \sum_{d=1}^{22} \epsilon^{{\rm pk},w}_{d}(E_\gamma).
\end{equation}
The values of $\epsilon^{{\rm pk},w}_{d}(E_\gamma)$ and $\epsilon^{\rm pk}$ decrease depending on $E_\gamma$ due to the loss of full absorption.

\begin{figure}[htbp]
\centering\includegraphics[width=0.5\linewidth,angle=0]{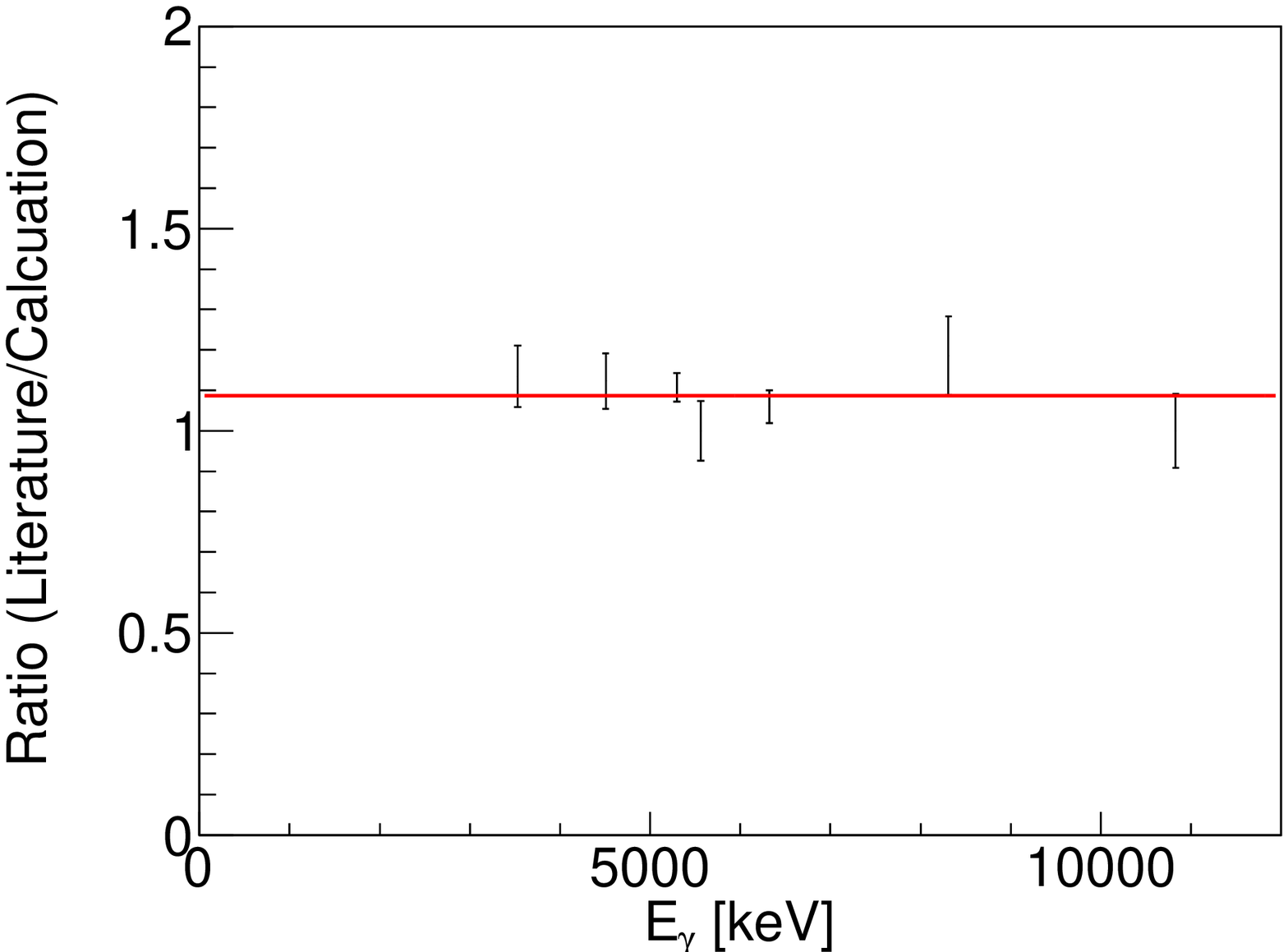}
\centering\includegraphics[width=0.5\linewidth,angle=0]{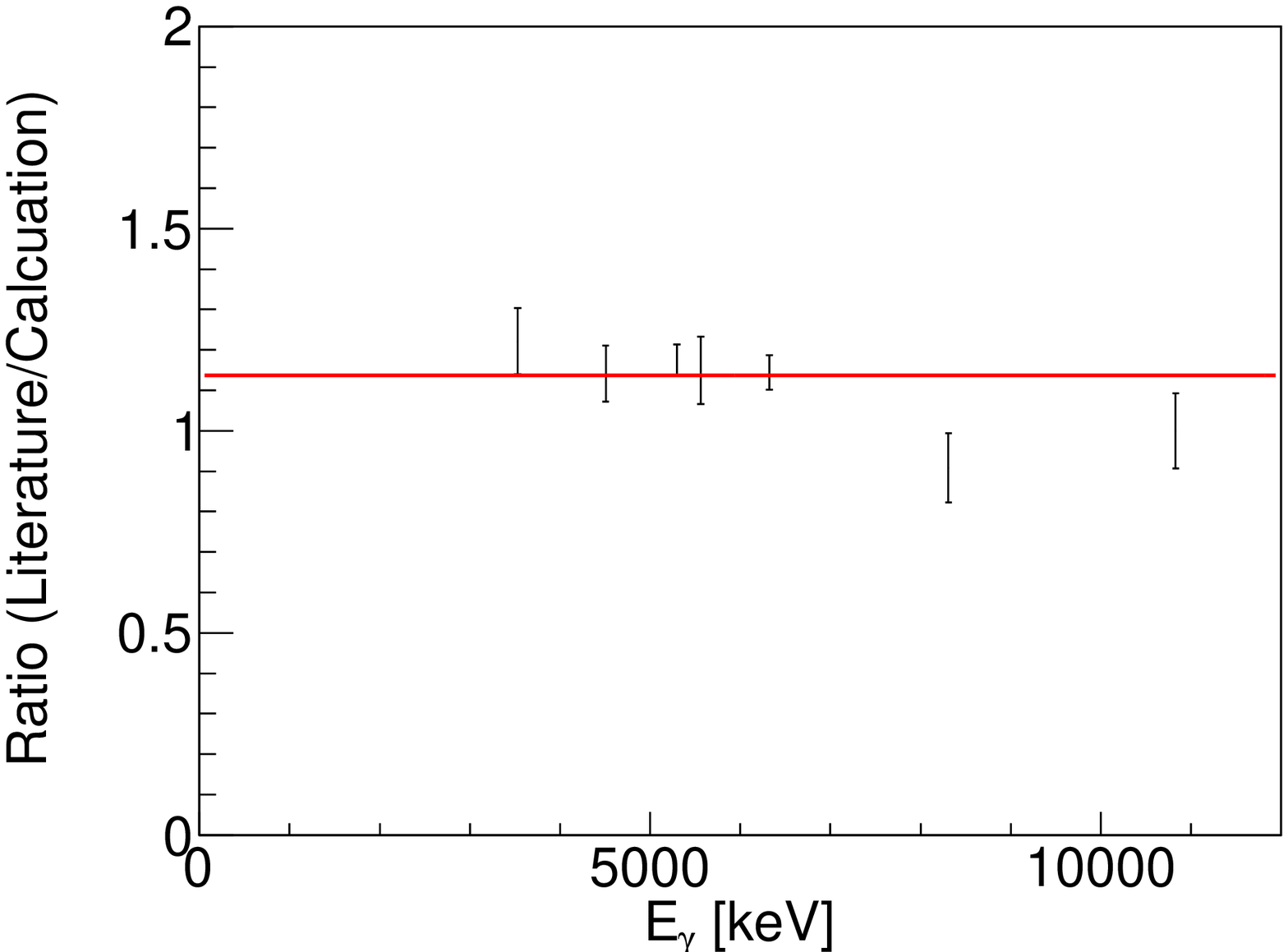}
\centering\includegraphics[width=0.5\linewidth,angle=0]{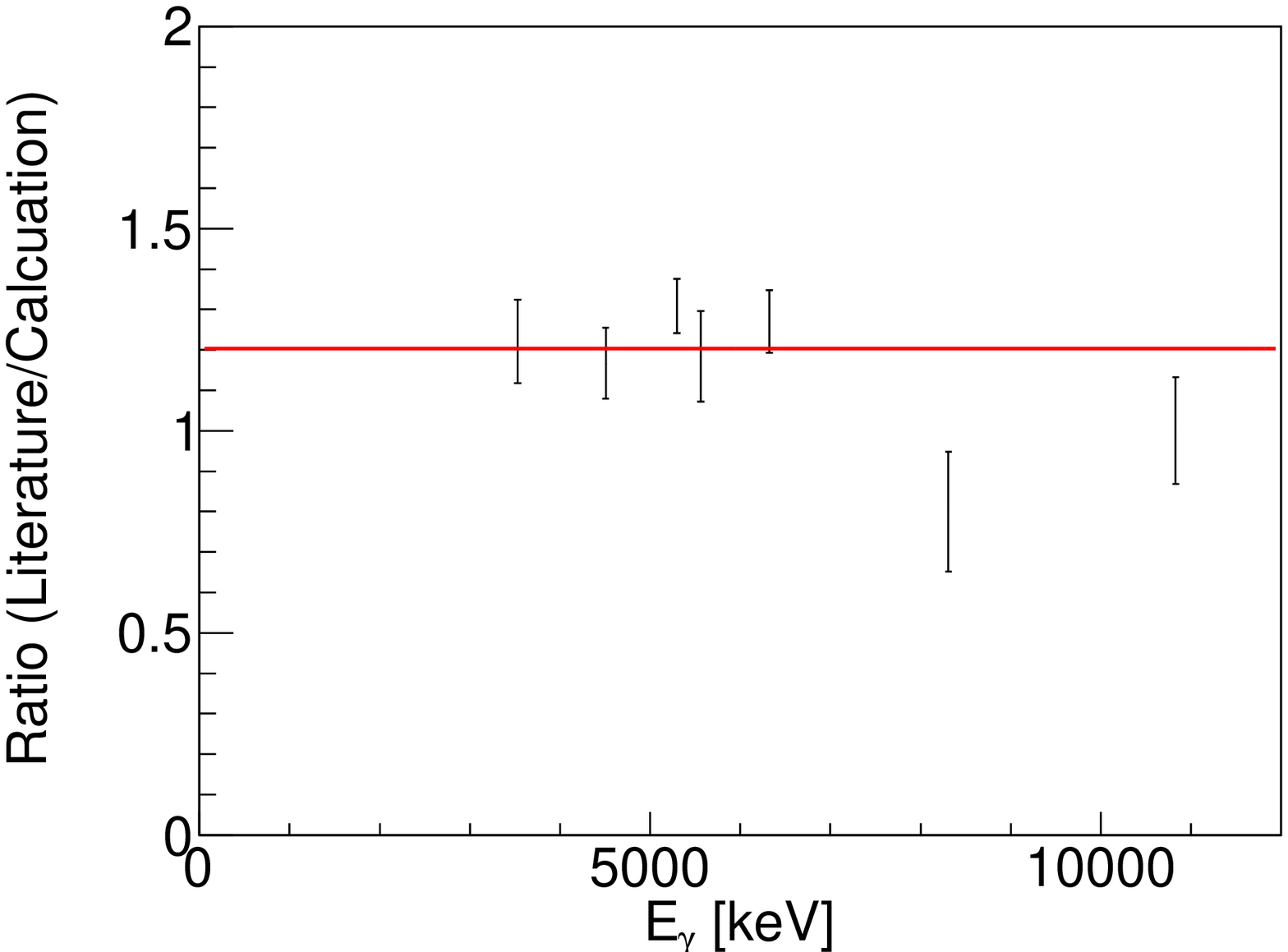}
\caption{\label{fig:14N_intensity}
Comparisons of the literature values~\cite{gamma_intensity} of the $\gamma$-ray intensity and the calculated values. Comparisons of detectors d8, d22, and d16 are shown in the top, middle, and bottom figures, respectively. The vertical axis is the ratio of intensities of the literature values and the calculated values of $\gamma$-rays from the $^{14}$N(n,$\gamma$) reaction. The horizontal axis is the energy of the $\gamma$-ray.
}
\end{figure}

Figure \ref{fig:14N_intensity} shows comparisons of the literature values~\cite{gamma_intensity} of the $\gamma$-rays intensities from the $^{14}$N(n,$\gamma$) reaction and our values, which are calculated by using measurements and the simulation.
The vertical axis is normalized to unity at 10829.1 keV.
The errors in Fig.~\ref{fig:14N_intensity} are given by the errors of literature values and the statistics from the measurement for the melamine target.
Our expectation was that the data points scatter linearly and can be fitted with a constant parameter, as represented by the solid line in Fig.~\ref{fig:14N_intensity}.
\textcolor{black}{The mean of the $\chi^2$/ndf of the fittings was about 1.5.
We were able to verify the simulation in the wide $\gamma$-ray energy range up to 11 MeV.}

Calculating Eq.\ref{eq:epsilon_peak} for the 661.7 keV peak of the $^{137}{\rm Cs}$ source, the difference of detection efficiency between measurement and simulation is about 10\% on average for all detectors.
\textcolor{black}{This, represented by a solid line in Fig. 16, deviated from unity, and the difference was about 8\% on average.}
The main reasons of the deviation of the absolute detection efficiency are the inaccuracy of the detector position, the difference in the size of germanium crystals due to crystal growth, and the difference in the sensible volume inside the crystals due to the difference in the cooling performances of each detector.
However, when the energy dependence is reproduced, it is possible to extrapolate an efficiency to arbitrary energy region by using radiation sources or targets which have not been observed angular distribution, such as $^{14}$N(n, $\gamma$).

\begin{landscape}
\begin{table}[htbp]
	\centering
	\small
	\begin{tabular}{|c || c | c | c | c | c | c | c | }
	\hline
	$E_\gamma=1\,[{\rm MeV}]$ & $\epsilon^{{\rm pk},w}_{d}(E_\gamma)$ & $\bar{P}_{d,1}$ & $\bar{P}_{d,2}$ & $\bar{P}_{d,3}$ & $\bar{P}_{d,4}$ & $\bar{P}_{d,5}$ & $\bar{P}_{d,6}$  \\ \hline
d1	 & 	$0.00115$	 & 	$0.000$	 & 	$-0.489$	 & 	$0.000$	 & 	$0.348$	 & 	$0.000$	 & 	$-0.266$ \\ \cline{1-8}
d2	 & 	$0.00101$	 & 	$0.000$	 & 	$-0.490$	 & 	$0.000$	 & 	$0.350$	 & 	$0.000$	 & 	$-0.270$ \\ \cline{1-8}
d3	 & 	$0.00105$	 & 	$0.302$	 & 	$-0.354$	 & 	$-0.370$	 & 	$0.062$	 & 	$0.310$	 & 	$0.118$ \\ \cline{1-8}
d4	 & 	$0.00104$	 & 	$0.302$	 & 	$-0.354$	 & 	$-0.370$	 & 	$0.062$	 & 	$0.310$	 & 	$0.118$ \\ \cline{1-8}
d5	 & 	$0.00103$	 & 	$0.000$	 & 	$-0.490$	 & 	$0.000$	 & 	$0.350$	 & 	$0.000$	 & 	$-0.270$ \\ \cline{1-8}
d6	 & 	$0.00105$	 & 	$-0.302$	 & 	$-0.354$	 & 	$0.370$	 & 	$0.062$	 & 	$-0.310$	 & 	$0.118$ \\ \cline{1-8}
d7	 & 	$0.00105$	 & 	$-0.302$	 & 	$-0.354$	 & 	$0.370$	 & 	$0.062$	 & 	$-0.310$	 & 	$0.118$ \\ \cline{1-8}
d8	 & 	$0.00115$	 & 	$0.000$	 & 	$-0.489$	 & 	$0.000$	 & 	$0.347$	 & 	$0.000$	 & 	$-0.265$ \\ \cline{1-8}
d9	 & 	$0.00101$	 & 	$0.000$	 & 	$-0.490$	 & 	$0.000$	 & 	$0.349$	 & 	$0.000$	 & 	$-0.269$ \\ \cline{1-8}
d10	 & 	$0.00105$	 & 	$0.302$	 & 	$-0.354$	 & 	$-0.370$	 & 	$0.062$	 & 	$0.310$	 & 	$0.118$ \\ \cline{1-8}
d11	 & 	$0.00103$	 & 	$0.302$	 & 	$-0.354$	 & 	$-0.370$	 & 	$0.062$	 & 	$0.310$	 & 	$0.118$ \\ \cline{1-8}
d12	 & 	$0.00101$	 & 	$0.000$	 & 	$-0.490$	 & 	$0.000$	 & 	$0.349$	 & 	$0.000$	 & 	$-0.269$ \\ \cline{1-8}
d13	 & 	$0.00105$	 & 	$-0.302$	 & 	$-0.354$	 & 	$0.370$	 & 	$0.062$	 & 	$-0.310$	 & 	$0.118$ \\ \cline{1-8}
d14	 & 	$0.00106$	 & 	$-0.302$	 & 	$-0.354$	 & 	$0.370$	 & 	$0.062$	 & 	$-0.310$	 & 	$0.118$ \\ \cline{1-8}
d15	 & 	$0.00108$	 & 	$-0.804$	 & 	$0.474$	 & 	$-0.109$	 & 	$-0.188$	 & 	$0.346$	 & 	$-0.347$ \\ \cline{1-8}
d16	 & 	$0.00053$	 & 	$-0.308$	 & 	$-0.352$	 & 	$0.379$	 & 	$0.054$	 & 	$-0.320$	 & 	$0.133$ \\ \cline{1-8}
d17	 & 	$0.00102$	 & 	$0.307$	 & 	$-0.349$	 & 	$-0.374$	 & 	$0.053$	 & 	$0.308$	 & 	$0.127$ \\ \cline{1-8}
d18	 & 	$0.00109$	 & 	$0.804$	 & 	$0.474$	 & 	$0.109$	 & 	$-0.188$	 & 	$-0.346$	 & 	$-0.347$ \\ \cline{1-8}
d19	 & 	$0.00108$	 & 	$0.804$	 & 	$0.474$	 & 	$0.109$	 & 	$-0.188$	 & 	$-0.346$	 & 	$-0.347$ \\ \cline{1-8}
d20	 & 	$0.00101$	 & 	$0.307$	 & 	$-0.349$	 & 	$-0.374$	 & 	$0.053$	 & 	$0.309$	 & 	$0.127$ \\ \cline{1-8}
d21	 & 	$0.00101$	 & 	$-0.307$	 & 	$-0.349$	 & 	$0.374$	 & 	$0.053$	 & 	$-0.309$	 & 	$0.127$ \\ \cline{1-8}
d22	 & 	$0.00108$	 & 	$-0.804$	 & 	$0.474$	 & 	$-0.109$	 & 	$-0.188$	 & 	$0.346$	 & 	$-0.347$ \\ \cline{1-8}
	\hline \hline
	 $\epsilon^{\rm pk} $	&	$0.02265$ & \multicolumn{6}{|c|}{} \\ \cline{1-8}
	\hline
	\end{tabular}
	\caption{Numerical values of $\bar{P}_{d,p}$ for $1 \le p \le 6$ for $E=1$ MeV.}
	\label{jikkou1MeV}
\end{table}
\end{landscape}
\begin{landscape}
\begin{table}[htbp]
	\centering
		\small
	\begin{tabular}{|c || c | c | c | c | c | c | c | }
	\hline
	$E_\gamma=2\,[{\rm MeV}]$ & $\epsilon^{{\rm pk},w}_{d}(E_\gamma)$ & $\bar{P}_{d,1}$ & $\bar{P}_{d,2}$ & $\bar{P}_{d,3}$ & $\bar{P}_{d,4}$ & $\bar{P}_{d,5}$ & $\bar{P}_{d,6}$  \\ \hline
d1	 & 	$0.00089$	 & 	$0.000$	 & 	$-0.489$	 & 	$0.000$	 & 	$0.349$	 & 	$0.000$	 & 	$-0.268$ \\ \cline{1-8}
d2	 & 	$0.00079$	 & 	$0.000$	 & 	$-0.490$	 & 	$0.000$	 & 	$0.351$	 & 	$0.000$	 & 	$-0.271$ \\ \cline{1-8}
d3	 & 	$0.00082$	 & 	$0.300$	 & 	$-0.356$	 & 	$-0.369$	 & 	$0.065$	 & 	$0.312$	 & 	$0.115$ \\ \cline{1-8}
d4	 & 	$0.00080$	 & 	$0.300$	 & 	$-0.356$	 & 	$-0.369$	 & 	$0.065$	 & 	$0.312$	 & 	$0.115$ \\ \cline{1-8}
d5	 & 	$0.00080$	 & 	$0.000$	 & 	$-0.490$	 & 	$0.000$	 & 	$0.351$	 & 	$0.000$	 & 	$-0.271$ \\ \cline{1-8}
d6	 & 	$0.00082$	 & 	$-0.300$	 & 	$-0.356$	 & 	$0.369$	 & 	$0.065$	 & 	$-0.312$	 & 	$0.115$ \\ \cline{1-8}
d7	 & 	$0.00082$	 & 	$-0.300$	 & 	$-0.356$	 & 	$0.369$	 & 	$0.065$	 & 	$-0.312$	 & 	$0.115$ \\ \cline{1-8}
d8	 & 	$0.00089$	 & 	$0.000$	 & 	$-0.489$	 & 	$0.000$	 & 	$0.348$	 & 	$0.000$	 & 	$-0.267$ \\ \cline{1-8}
d9	 & 	$0.00080$	 & 	$0.000$	 & 	$-0.490$	 & 	$0.000$	 & 	$0.351$	 & 	$0.000$	 & 	$-0.271$ \\ \cline{1-8}
d10	 & 	$0.00082$	 & 	$0.300$	 & 	$-0.356$	 & 	$-0.369$	 & 	$0.065$	 & 	$0.312$	 & 	$0.115$ \\ \cline{1-8}
d11	 & 	$0.00081$	 & 	$0.300$	 & 	$-0.356$	 & 	$-0.369$	 & 	$0.065$	 & 	$0.312$	 & 	$0.115$ \\ \cline{1-8}
d12	 & 	$0.00080$	 & 	$0.000$	 & 	$-0.490$	 & 	$0.000$	 & 	$0.351$	 & 	$0.000$	 & 	$-0.271$ \\ \cline{1-8}
d13	 & 	$0.00082$	 & 	$-0.300$	 & 	$-0.356$	 & 	$0.369$	 & 	$0.065$	 & 	$-0.312$	 & 	$0.115$ \\ \cline{1-8}
d14	 & 	$0.00082$	 & 	$-0.300$	 & 	$-0.356$	 & 	$0.369$	 & 	$0.065$	 & 	$-0.312$	 & 	$0.116$ \\ \cline{1-8}
d15	 & 	$0.00087$	 & 	$-0.804$	 & 	$0.474$	 & 	$-0.108$	 & 	$-0.189$	 & 	$0.347$	 & 	$-0.348$ \\ \cline{1-8}
d16	 & 	$0.00041$	 & 	$-0.308$	 & 	$-0.352$	 & 	$0.379$	 & 	$0.054$	 & 	$-0.321$	 & 	$0.134$ \\ \cline{1-8}
d17	 & 	$0.00083$	 & 	$0.307$	 & 	$-0.349$	 & 	$-0.374$	 & 	$0.053$	 & 	$0.309$	 & 	$0.127$ \\ \cline{1-8}
d18	 & 	$0.00087$	 & 	$0.804$	 & 	$0.474$	 & 	$0.108$	 & 	$-0.189$	 & 	$-0.347$	 & 	$-0.348$ \\ \cline{1-8}
d19	 & 	$0.00086$	 & 	$0.804$	 & 	$0.473$	 & 	$0.108$	 & 	$-0.189$	 & 	$-0.347$	 & 	$-0.348$ \\ \cline{1-8}
d20	 & 	$0.00083$	 & 	$0.307$	 & 	$-0.349$	 & 	$-0.374$	 & 	$0.053$	 & 	$0.309$	 & 	$0.128$ \\ \cline{1-8}
d21	 & 	$0.00083$	 & 	$-0.307$	 & 	$-0.349$	 & 	$0.374$	 & 	$0.052$	 & 	$-0.309$	 & 	$0.128$ \\ \cline{1-8}
d22	 & 	$0.00088$	 & 	$-0.804$	 & 	$0.474$	 & 	$-0.108$	 & 	$-0.189$	 & 	$0.347$	 & 	$-0.348$ \\ \cline{1-8}
	\hline \hline
	$\epsilon^{\rm pk} $	&	$0.01789$ & \multicolumn{6}{|c|}{} \\ \cline{1-8}
	\hline
	\end{tabular}
	\caption{Numerical values of $\bar{P}_{d,p}$ for $1 \le p \le 6$ for $E=2$ MeV.}
	\label{jikkou2MeV}
\end{table}
\end{landscape}
\begin{landscape}
\begin{table}[htbp]
	\centering
		\small
	\begin{tabular}{|c || c | c | c | c | c | c | c | }
	\hline
	$E_\gamma=3\,[{\rm MeV}]$ & $\epsilon^{{\rm pk},w}_{d}(E_\gamma)$ & $\bar{P}_{d,1}$ & $\bar{P}_{d,2}$ & $\bar{P}_{d,3}$ & $\bar{P}_{d,4}$ & $\bar{P}_{d,5}$ & $\bar{P}_{d,6}$  \\ \hline
d1	 & 	$0.00070$	 & 	$0.000$	 & 	$-0.490$	 & 	$0.000$	 & 	$0.349$	 & 	$0.000$	 & 	$-0.269$ \\ \cline{1-8}
d2	 & 	$0.00063$	 & 	$0.000$	 & 	$-0.490$	 & 	$0.000$	 & 	$0.352$	 & 	$0.000$	 & 	$-0.273$ \\ \cline{1-8}
d3	 & 	$0.00065$	 & 	$0.299$	 & 	$-0.357$	 & 	$-0.369$	 & 	$0.066$	 & 	$0.313$	 & 	$0.115$ \\ \cline{1-8}
d4	 & 	$0.00063$	 & 	$0.299$	 & 	$-0.357$	 & 	$-0.369$	 & 	$0.066$	 & 	$0.313$	 & 	$0.114$ \\ \cline{1-8}
d5	 & 	$0.00063$	 & 	$0.000$	 & 	$-0.491$	 & 	$0.000$	 & 	$0.352$	 & 	$0.000$	 & 	$-0.273$ \\ \cline{1-8}
d6	 & 	$0.00065$	 & 	$-0.300$	 & 	$-0.357$	 & 	$0.369$	 & 	$0.066$	 & 	$-0.313$	 & 	$0.115$ \\ \cline{1-8}
d7	 & 	$0.00065$	 & 	$-0.299$	 & 	$-0.357$	 & 	$0.369$	 & 	$0.066$	 & 	$-0.313$	 & 	$0.115$ \\ \cline{1-8}
d8	 & 	$0.00070$	 & 	$0.000$	 & 	$-0.490$	 & 	$0.000$	 & 	$0.349$	 & 	$0.000$	 & 	$-0.269$ \\ \cline{1-8}
d9	 & 	$0.00064$	 & 	$0.000$	 & 	$-0.490$	 & 	$0.000$	 & 	$0.351$	 & 	$0.000$	 & 	$-0.272$ \\ \cline{1-8}
d10	 & 	$0.00065$	 & 	$0.299$	 & 	$-0.357$	 & 	$-0.369$	 & 	$0.066$	 & 	$0.313$	 & 	$0.115$ \\ \cline{1-8}
d11	 & 	$0.00064$	 & 	$0.300$	 & 	$-0.357$	 & 	$-0.369$	 & 	$0.066$	 & 	$0.313$	 & 	$0.115$ \\ \cline{1-8}
d12	 & 	$0.00064$	 & 	$0.000$	 & 	$-0.490$	 & 	$0.000$	 & 	$0.351$	 & 	$0.000$	 & 	$-0.272$ \\ \cline{1-8}
d13	 & 	$0.00065$	 & 	$-0.299$	 & 	$-0.357$	 & 	$0.369$	 & 	$0.066$	 & 	$-0.313$	 & 	$0.115$ \\ \cline{1-8}
d14	 & 	$0.00065$	 & 	$-0.300$	 & 	$-0.357$	 & 	$0.369$	 & 	$0.066$	 & 	$-0.313$	 & 	$0.115$ \\ \cline{1-8}
d15	 & 	$0.00069$	 & 	$-0.804$	 & 	$0.473$	 & 	$-0.108$	 & 	$-0.190$	 & 	$0.348$	 & 	$-0.349$ \\ \cline{1-8}
d16	 & 	$0.00032$	 & 	$-0.307$	 & 	$-0.352$	 & 	$0.379$	 & 	$0.055$	 & 	$-0.321$	 & 	$0.134$ \\ \cline{1-8}
d17	 & 	$0.00068$	 & 	$0.307$	 & 	$-0.350$	 & 	$-0.374$	 & 	$0.053$	 & 	$0.310$	 & 	$0.128$ \\ \cline{1-8}
d18	 & 	$0.00070$	 & 	$0.804$	 & 	$0.473$	 & 	$0.108$	 & 	$-0.190$	 & 	$-0.348$	 & 	$-0.349$ \\ \cline{1-8}
d19	 & 	$0.00070$	 & 	$0.804$	 & 	$0.473$	 & 	$0.107$	 & 	$-0.190$	 & 	$-0.349$	 & 	$-0.349$ \\ \cline{1-8}
d20	 & 	$0.00068$	 & 	$0.307$	 & 	$-0.350$	 & 	$-0.374$	 & 	$0.053$	 & 	$0.310$	 & 	$0.128$ \\ \cline{1-8}
d21	 & 	$0.00068$	 & 	$-0.307$	 & 	$-0.349$	 & 	$0.375$	 & 	$0.053$	 & 	$-0.310$	 & 	$0.128$ \\ \cline{1-8}
d22	 & 	$0.00070$	 & 	$-0.804$	 & 	$0.474$	 & 	$-0.108$	 & 	$-0.190$	 & 	$0.348$	 & 	$-0.349$ \\ \cline{1-8}
	\hline \hline
	$\epsilon^{\rm pk} $	&	$0.01426$ & \multicolumn{6}{|c|}{} \\ \cline{1-8}
	\hline
	\end{tabular}
	\caption{Numerical values of $\bar{P}_{d,p}$ for $1 \le p \le 6$ for $E=3$ MeV.}
	\label{jikkou3MeV}
\end{table}
\end{landscape}
\begin{landscape}
\begin{table}[htbp]
	\centering
		\small
	\begin{tabular}{|c || c | c | c | c | c | c | c | }
	\hline
	$E_\gamma=4\,[{\rm MeV}]$ & $\epsilon^{{\rm pk},w}_{d}(E_\gamma)$ & $\bar{P}_{d,1}$ & $\bar{P}_{d,2}$ & $\bar{P}_{d,3}$ & $\bar{P}_{d,4}$ & $\bar{P}_{d,5}$ & $\bar{P}_{d,6}$  \\ \hline
d1	 & 	$0.00056$	 & 	$0.000$	 & 	$-0.490$	 & 	$0.000$	 & 	$0.350$	 & 	$0.000$	 & 	$-0.271$ \\ \cline{1-8}
d2	 & 	$0.00051$	 & 	$0.000$	 & 	$-0.491$	 & 	$0.000$	 & 	$0.352$	 & 	$0.000$	 & 	$-0.274$ \\ \cline{1-8}
d3	 & 	$0.00052$	 & 	$0.299$	 & 	$-0.357$	 & 	$-0.369$	 & 	$0.067$	 & 	$0.314$	 & 	$0.114$ \\ \cline{1-8}
d4	 & 	$0.00052$	 & 	$0.299$	 & 	$-0.357$	 & 	$-0.369$	 & 	$0.067$	 & 	$0.314$	 & 	$0.114$ \\ \cline{1-8}
d5	 & 	$0.00051$	 & 	$0.000$	 & 	$-0.491$	 & 	$0.000$	 & 	$0.352$	 & 	$0.000$	 & 	$-0.274$ \\ \cline{1-8}
d6	 & 	$0.00053$	 & 	$-0.299$	 & 	$-0.357$	 & 	$0.369$	 & 	$0.067$	 & 	$-0.314$	 & 	$0.114$ \\ \cline{1-8}
d7	 & 	$0.00053$	 & 	$-0.299$	 & 	$-0.357$	 & 	$0.369$	 & 	$0.067$	 & 	$-0.314$	 & 	$0.114$ \\ \cline{1-8}
d8	 & 	$0.00056$	 & 	$0.000$	 & 	$-0.490$	 & 	$0.000$	 & 	$0.350$	 & 	$0.000$	 & 	$-0.270$ \\ \cline{1-8}
d9	 & 	$0.00051$	 & 	$0.000$	 & 	$-0.491$	 & 	$0.000$	 & 	$0.352$	 & 	$0.000$	 & 	$-0.273$ \\ \cline{1-8}
d10	 & 	$0.00053$	 & 	$0.299$	 & 	$-0.357$	 & 	$-0.369$	 & 	$0.067$	 & 	$0.314$	 & 	$0.115$ \\ \cline{1-8}
d11	 & 	$0.00051$	 & 	$0.299$	 & 	$-0.357$	 & 	$-0.369$	 & 	$0.067$	 & 	$0.314$	 & 	$0.115$ \\ \cline{1-8}
d12	 & 	$0.00051$	 & 	$0.000$	 & 	$-0.491$	 & 	$0.000$	 & 	$0.352$	 & 	$0.000$	 & 	$-0.273$ \\ \cline{1-8}
d13	 & 	$0.00053$	 & 	$-0.299$	 & 	$-0.357$	 & 	$0.369$	 & 	$0.067$	 & 	$-0.314$	 & 	$0.114$ \\ \cline{1-8}
d14	 & 	$0.00053$	 & 	$-0.299$	 & 	$-0.357$	 & 	$0.369$	 & 	$0.067$	 & 	$-0.314$	 & 	$0.114$ \\ \cline{1-8}
d15	 & 	$0.00057$	 & 	$-0.804$	 & 	$0.473$	 & 	$-0.108$	 & 	$-0.190$	 & 	$0.349$	 & 	$-0.350$ \\ \cline{1-8}
d16	 & 	$0.00025$	 & 	$-0.308$	 & 	$-0.352$	 & 	$0.380$	 & 	$0.054$	 & 	$-0.322$	 & 	$0.135$ \\ \cline{1-8}
d17	 & 	$0.00056$	 & 	$0.307$	 & 	$-0.350$	 & 	$-0.375$	 & 	$0.053$	 & 	$0.311$	 & 	$0.128$ \\ \cline{1-8}
d18	 & 	$0.00057$	 & 	$0.804$	 & 	$0.473$	 & 	$0.108$	 & 	$-0.191$	 & 	$-0.349$	 & 	$-0.350$ \\ \cline{1-8}
d19	 & 	$0.00057$	 & 	$0.804$	 & 	$0.473$	 & 	$0.107$	 & 	$-0.191$	 & 	$-0.349$	 & 	$-0.350$ \\ \cline{1-8}
d20	 & 	$0.00056$	 & 	$0.307$	 & 	$-0.350$	 & 	$-0.375$	 & 	$0.053$	 & 	$0.311$	 & 	$0.128$ \\ \cline{1-8}
d21	 & 	$0.00056$	 & 	$-0.307$	 & 	$-0.350$	 & 	$0.375$	 & 	$0.053$	 & 	$-0.311$	 & 	$0.128$ \\ \cline{1-8}
d22	 & 	$0.00057$	 & 	$-0.804$	 & 	$0.473$	 & 	$-0.107$	 & 	$-0.191$	 & 	$0.350$	 & 	$-0.350$ \\ \cline{1-8}
	\hline \hline
	$\epsilon^{\rm pk} $	&	$0.01156$ & \multicolumn{6}{|c|}{} \\ \cline{1-8}
	\hline
	\end{tabular}
	\caption{Numerical values of $\bar{P}_{d,p}$ for $1 \le p \le 6$ for $E=4$ MeV.}
	\label{jikkou4MeV}
\end{table}
\end{landscape}
\begin{landscape}
\begin{table}[htbp]
	\centering
		\small
	\begin{tabular}{|c || c | c | c | c | c | c | c | }
	\hline
	$E_\gamma=5\,[{\rm MeV}]$ & $\epsilon^{{\rm pk},w}_{d}(E_\gamma)$ & $\bar{P}_{d,1}$ & $\bar{P}_{d,2}$ & $\bar{P}_{d,3}$ & $\bar{P}_{d,4}$ & $\bar{P}_{d,5}$ & $\bar{P}_{d,6}$  \\ \hline
d1	 & 	$0.00046$	 & 	$0.000$	 & 	$-0.490$	 & 	$0.000$	 & 	$0.351$	 & 	$0.000$	 & 	$-0.272$ \\ \cline{1-8}
d2	 & 	$0.00042$	 & 	$0.000$	 & 	$-0.491$	 & 	$0.000$	 & 	$0.353$	 & 	$0.000$	 & 	$-0.275$ \\ \cline{1-8}
d3	 & 	$0.00043$	 & 	$0.299$	 & 	$-0.358$	 & 	$-0.369$	 & 	$0.068$	 & 	$0.315$	 & 	$0.114$ \\ \cline{1-8}
d4	 & 	$0.00042$	 & 	$0.299$	 & 	$-0.358$	 & 	$-0.369$	 & 	$0.068$	 & 	$0.315$	 & 	$0.114$ \\ \cline{1-8}
d5	 & 	$0.00042$	 & 	$0.000$	 & 	$-0.491$	 & 	$0.000$	 & 	$0.353$	 & 	$0.000$	 & 	$-0.275$ \\ \cline{1-8}
d6	 & 	$0.00043$	 & 	$-0.299$	 & 	$-0.358$	 & 	$0.369$	 & 	$0.068$	 & 	$-0.315$	 & 	$0.114$ \\ \cline{1-8}
d7	 & 	$0.00043$	 & 	$-0.299$	 & 	$-0.358$	 & 	$0.369$	 & 	$0.068$	 & 	$-0.315$	 & 	$0.114$ \\ \cline{1-8}
d8	 & 	$0.00046$	 & 	$0.000$	 & 	$-0.490$	 & 	$0.000$	 & 	$0.351$	 & 	$0.000$	 & 	$-0.272$ \\ \cline{1-8}
d9	 & 	$0.00043$	 & 	$0.000$	 & 	$-0.491$	 & 	$0.000$	 & 	$0.353$	 & 	$0.000$	 & 	$-0.275$ \\ \cline{1-8}
d10	 & 	$0.00043$	 & 	$0.299$	 & 	$-0.358$	 & 	$-0.370$	 & 	$0.067$	 & 	$0.315$	 & 	$0.115$ \\ \cline{1-8}
d11	 & 	$0.00042$	 & 	$0.299$	 & 	$-0.358$	 & 	$-0.369$	 & 	$0.068$	 & 	$0.315$	 & 	$0.114$ \\ \cline{1-8}
d12	 & 	$0.00042$	 & 	$0.000$	 & 	$-0.491$	 & 	$0.000$	 & 	$0.353$	 & 	$0.000$	 & 	$-0.275$ \\ \cline{1-8}
d13	 & 	$0.00043$	 & 	$-0.299$	 & 	$-0.358$	 & 	$0.369$	 & 	$0.068$	 & 	$-0.315$	 & 	$0.114$ \\ \cline{1-8}
d14	 & 	$0.00043$	 & 	$-0.299$	 & 	$-0.358$	 & 	$0.370$	 & 	$0.067$	 & 	$-0.315$	 & 	$0.115$ \\ \cline{1-8}
d15	 & 	$0.00047$	 & 	$-0.804$	 & 	$0.473$	 & 	$-0.107$	 & 	$-0.191$	 & 	$0.350$	 & 	$-0.351$ \\ \cline{1-8}
d16	 & 	$0.00020$	 & 	$-0.308$	 & 	$-0.352$	 & 	$0.380$	 & 	$0.055$	 & 	$-0.322$	 & 	$0.134$ \\ \cline{1-8}
d17	 & 	$0.00046$	 & 	$0.307$	 & 	$-0.350$	 & 	$-0.375$	 & 	$0.053$	 & 	$0.312$	 & 	$0.129$ \\ \cline{1-8}
d18	 & 	$0.00047$	 & 	$0.804$	 & 	$0.473$	 & 	$0.107$	 & 	$-0.191$	 & 	$-0.350$	 & 	$-0.351$ \\ \cline{1-8}
d19	 & 	$0.00047$	 & 	$0.804$	 & 	$0.473$	 & 	$0.107$	 & 	$-0.191$	 & 	$-0.350$	 & 	$-0.351$ \\ \cline{1-8}
d20	 & 	$0.00046$	 & 	$0.307$	 & 	$-0.350$	 & 	$-0.375$	 & 	$0.054$	 & 	$0.312$	 & 	$0.128$ \\ \cline{1-8}
d21	 & 	$0.00046$	 & 	$-0.307$	 & 	$-0.350$	 & 	$0.375$	 & 	$0.053$	 & 	$-0.312$	 & 	$0.128$ \\ \cline{1-8}
d22	 & 	$0.00047$	 & 	$-0.804$	 & 	$0.474$	 & 	$-0.107$	 & 	$-0.191$	 & 	$0.350$	 & 	$-0.352$ \\ \cline{1-8}
	\hline \hline
	$\epsilon^{\rm pk} $	&	$0.00951$ & \multicolumn{6}{|c|}{} \\ \cline{1-8}
	\hline
	\end{tabular}
	\caption{Numerical values of $\bar{P}_{d,p}$ for $1 \le p \le 6$ for $E=5$ MeV.}
	\label{jikkou5MeV}
\end{table}
\end{landscape}
\begin{landscape}
\begin{table}[htbp]
	\centering
		\small
	\begin{tabular}{|c || c | c | c | c | c | c | c | }
	\hline
	$E_\gamma=6\,[{\rm MeV}]$ & $\epsilon^{{\rm pk},w}_{d}(E_\gamma)$ & $\bar{P}_{d,1}$ & $\bar{P}_{d,2}$ & $\bar{P}_{d,3}$ & $\bar{P}_{d,4}$ & $\bar{P}_{d,5}$ & $\bar{P}_{d,6}$  \\ \hline
d1	 & 	$0.00039$	 & 	$0.000$	 & 	$-0.491$	 & 	$0.000$	 & 	$0.352$	 & 	$0.000$	 & 	$-0.273$ \\ \cline{1-8}
d2	 & 	$0.00035$	 & 	$0.000$	 & 	$-0.491$	 & 	$0.000$	 & 	$0.354$	 & 	$0.000$	 & 	$-0.276$ \\ \cline{1-8}
d3	 & 	$0.00036$	 & 	$0.299$	 & 	$-0.358$	 & 	$-0.369$	 & 	$0.068$	 & 	$0.316$	 & 	$0.114$ \\ \cline{1-8}
d4	 & 	$0.00035$	 & 	$0.299$	 & 	$-0.358$	 & 	$-0.369$	 & 	$0.068$	 & 	$0.316$	 & 	$0.114$ \\ \cline{1-8}
d5	 & 	$0.00035$	 & 	$0.000$	 & 	$-0.491$	 & 	$0.000$	 & 	$0.354$	 & 	$0.000$	 & 	$-0.276$ \\ \cline{1-8}
d6	 & 	$0.00036$	 & 	$-0.298$	 & 	$-0.358$	 & 	$0.369$	 & 	$0.069$	 & 	$-0.316$	 & 	$0.114$ \\ \cline{1-8}
d7	 & 	$0.00036$	 & 	$-0.299$	 & 	$-0.358$	 & 	$0.370$	 & 	$0.068$	 & 	$-0.316$	 & 	$0.115$ \\ \cline{1-8}
d8	 & 	$0.00039$	 & 	$0.000$	 & 	$-0.491$	 & 	$0.000$	 & 	$0.352$	 & 	$0.000$	 & 	$-0.273$ \\ \cline{1-8}
d9	 & 	$0.00035$	 & 	$0.000$	 & 	$-0.491$	 & 	$0.000$	 & 	$0.354$	 & 	$0.000$	 & 	$-0.276$ \\ \cline{1-8}
d10	 & 	$0.00036$	 & 	$0.299$	 & 	$-0.358$	 & 	$-0.370$	 & 	$0.068$	 & 	$0.316$	 & 	$0.115$ \\ \cline{1-8}
d11	 & 	$0.00035$	 & 	$0.299$	 & 	$-0.358$	 & 	$-0.370$	 & 	$0.068$	 & 	$0.316$	 & 	$0.114$ \\ \cline{1-8}
d12	 & 	$0.00035$	 & 	$0.000$	 & 	$-0.491$	 & 	$0.000$	 & 	$0.353$	 & 	$0.000$	 & 	$-0.276$ \\ \cline{1-8}
d13	 & 	$0.00036$	 & 	$-0.299$	 & 	$-0.358$	 & 	$0.370$	 & 	$0.068$	 & 	$-0.316$	 & 	$0.115$ \\ \cline{1-8}
d14	 & 	$0.00036$	 & 	$-0.299$	 & 	$-0.358$	 & 	$0.370$	 & 	$0.068$	 & 	$-0.316$	 & 	$0.115$ \\ \cline{1-8}
d15	 & 	$0.00039$	 & 	$-0.804$	 & 	$0.474$	 & 	$-0.107$	 & 	$-0.191$	 & 	$0.351$	 & 	$-0.353$ \\ \cline{1-8}
d16	 & 	$0.00017$	 & 	$-0.308$	 & 	$-0.352$	 & 	$0.380$	 & 	$0.054$	 & 	$-0.323$	 & 	$0.135$ \\ \cline{1-8}
d17	 & 	$0.00039$	 & 	$0.307$	 & 	$-0.350$	 & 	$-0.375$	 & 	$0.053$	 & 	$0.312$	 & 	$0.129$ \\ \cline{1-8}
d18	 & 	$0.00040$	 & 	$0.804$	 & 	$0.473$	 & 	$0.107$	 & 	$-0.192$	 & 	$-0.351$	 & 	$-0.352$ \\ \cline{1-8}
d19	 & 	$0.00039$	 & 	$0.804$	 & 	$0.473$	 & 	$0.107$	 & 	$-0.192$	 & 	$-0.351$	 & 	$-0.353$ \\ \cline{1-8}
d20	 & 	$0.00039$	 & 	$0.307$	 & 	$-0.350$	 & 	$-0.376$	 & 	$0.053$	 & 	$0.312$	 & 	$0.129$ \\ \cline{1-8}
d21	 & 	$0.00039$	 & 	$-0.307$	 & 	$-0.350$	 & 	$0.375$	 & 	$0.053$	 & 	$-0.312$	 & 	$0.129$ \\ \cline{1-8}
d22	 & 	$0.00039$	 & 	$-0.804$	 & 	$0.473$	 & 	$-0.106$	 & 	$-0.192$	 & 	$0.352$	 & 	$-0.353$ \\ \cline{1-8}
	\hline \hline
	$\epsilon^{\rm pk} $	&	$0.00794$ & \multicolumn{6}{|c|}{} \\ \cline{1-8}
	\hline
	\end{tabular}
	\caption{Numerical values of $\bar{P}_{d,p}$ for $1 \le p \le 6$ for $E=6$ MeV.}
	\label{jikkou6MeV}
\end{table}
\end{landscape}
\begin{landscape}
\begin{table}[htbp]
	\centering
		\small
	\begin{tabular}{|c || c | c | c | c | c | c | c | }
	\hline
	$E_\gamma=7\,[{\rm MeV}]$ & $\epsilon^{{\rm pk},w}_{d}(E_\gamma)$ & $\bar{P}_{d,1}$ & $\bar{P}_{d,2}$ & $\bar{P}_{d,3}$ & $\bar{P}_{d,4}$ & $\bar{P}_{d,5}$ & $\bar{P}_{d,6}$  \\ \hline
d1	 & 	$0.00032$	 & 	$0.000$	 & 	$-0.491$	 & 	$0.000$	 & 	$0.352$	 & 	$0.000$	 & 	$-0.274$ \\ \cline{1-8}
d2	 & 	$0.00030$	 & 	$0.000$	 & 	$-0.492$	 & 	$0.000$	 & 	$0.354$	 & 	$0.000$	 & 	$-0.277$ \\ \cline{1-8}
d3	 & 	$0.00030$	 & 	$0.299$	 & 	$-0.359$	 & 	$-0.370$	 & 	$0.069$	 & 	$0.317$	 & 	$0.115$ \\ \cline{1-8}
d4	 & 	$0.00029$	 & 	$0.299$	 & 	$-0.358$	 & 	$-0.370$	 & 	$0.068$	 & 	$0.317$	 & 	$0.115$ \\ \cline{1-8}
d5	 & 	$0.00029$	 & 	$0.000$	 & 	$-0.492$	 & 	$0.000$	 & 	$0.354$	 & 	$0.000$	 & 	$-0.277$ \\ \cline{1-8}
d6	 & 	$0.00030$	 & 	$-0.299$	 & 	$-0.358$	 & 	$0.370$	 & 	$0.068$	 & 	$-0.317$	 & 	$0.115$ \\ \cline{1-8}
d7	 & 	$0.00030$	 & 	$-0.299$	 & 	$-0.359$	 & 	$0.370$	 & 	$0.069$	 & 	$-0.317$	 & 	$0.115$ \\ \cline{1-8}
d8	 & 	$0.00032$	 & 	$0.000$	 & 	$-0.491$	 & 	$0.000$	 & 	$0.352$	 & 	$0.000$	 & 	$-0.273$ \\ \cline{1-8}
d9	 & 	$0.00030$	 & 	$0.000$	 & 	$-0.492$	 & 	$0.000$	 & 	$0.354$	 & 	$0.000$	 & 	$-0.277$ \\ \cline{1-8}
d10	 & 	$0.00030$	 & 	$0.299$	 & 	$-0.358$	 & 	$-0.370$	 & 	$0.068$	 & 	$0.317$	 & 	$0.115$ \\ \cline{1-8}
d11	 & 	$0.00029$	 & 	$0.299$	 & 	$-0.359$	 & 	$-0.370$	 & 	$0.069$	 & 	$0.317$	 & 	$0.114$ \\ \cline{1-8}
d12	 & 	$0.00029$	 & 	$0.000$	 & 	$-0.492$	 & 	$0.000$	 & 	$0.354$	 & 	$0.000$	 & 	$-0.277$ \\ \cline{1-8}
d13	 & 	$0.00030$	 & 	$-0.299$	 & 	$-0.358$	 & 	$0.370$	 & 	$0.068$	 & 	$-0.317$	 & 	$0.115$ \\ \cline{1-8}
d14	 & 	$0.00030$	 & 	$-0.299$	 & 	$-0.359$	 & 	$0.370$	 & 	$0.069$	 & 	$-0.317$	 & 	$0.114$ \\ \cline{1-8}
d15	 & 	$0.00033$	 & 	$-0.804$	 & 	$0.474$	 & 	$-0.107$	 & 	$-0.192$	 & 	$0.352$	 & 	$-0.354$ \\ \cline{1-8}
d16	 & 	$0.00014$	 & 	$-0.308$	 & 	$-0.352$	 & 	$0.381$	 & 	$0.054$	 & 	$-0.323$	 & 	$0.135$ \\ \cline{1-8}
d17	 & 	$0.00033$	 & 	$0.307$	 & 	$-0.350$	 & 	$-0.376$	 & 	$0.053$	 & 	$0.313$	 & 	$0.129$ \\ \cline{1-8}
d18	 & 	$0.00033$	 & 	$0.804$	 & 	$0.474$	 & 	$0.107$	 & 	$-0.192$	 & 	$-0.352$	 & 	$-0.354$ \\ \cline{1-8}
d19	 & 	$0.00033$	 & 	$0.804$	 & 	$0.474$	 & 	$0.107$	 & 	$-0.192$	 & 	$-0.352$	 & 	$-0.354$ \\ \cline{1-8}
d20	 & 	$0.00033$	 & 	$0.307$	 & 	$-0.350$	 & 	$-0.376$	 & 	$0.054$	 & 	$0.313$	 & 	$0.129$ \\ \cline{1-8}
d21	 & 	$0.00033$	 & 	$-0.307$	 & 	$-0.351$	 & 	$0.375$	 & 	$0.054$	 & 	$-0.313$	 & 	$0.129$ \\ \cline{1-8}
d22	 & 	$0.00033$	 & 	$-0.804$	 & 	$0.474$	 & 	$-0.107$	 & 	$-0.192$	 & 	$0.352$	 & 	$-0.354$ \\ \cline{1-8}
	\hline \hline
	$\epsilon^{\rm pk} $	&	$0.00666$ & \multicolumn{6}{|c|}{} \\ \cline{1-8}
	\hline
	\end{tabular}
	\caption{Numerical values of $\bar{P}_{d,p}$ for $1 \le p \le 6$ for $E=7$ MeV.}
	\label{jikkou7MeV}
\end{table}
\end{landscape}
\begin{landscape}
\begin{table}[htbp]
	\centering
		\small
	\begin{tabular}{|c || c | c | c | c | c | c | c | }
	\hline
	$E_\gamma=8\,[{\rm MeV}]$ & $\epsilon^{{\rm pk},w}_{d}(E_\gamma)$ & $\bar{P}_{d,1}$ & $\bar{P}_{d,2}$ & $\bar{P}_{d,3}$ & $\bar{P}_{d,4}$ & $\bar{P}_{d,5}$ & $\bar{P}_{d,6}$  \\ \hline
d1	 & 	$0.00027$	 & 	$0.000$	 & 	$-0.491$	 & 	$0.000$	 & 	$0.353$	 & 	$0.000$	 & 	$-0.275$ \\ \cline{1-8}
d2	 & 	$0.00025$	 & 	$0.000$	 & 	$-0.492$	 & 	$0.000$	 & 	$0.355$	 & 	$0.000$	 & 	$-0.278$ \\ \cline{1-8}
d3	 & 	$0.00025$	 & 	$0.299$	 & 	$-0.359$	 & 	$-0.370$	 & 	$0.069$	 & 	$0.317$	 & 	$0.115$ \\ \cline{1-8}
d4	 & 	$0.00025$	 & 	$0.298$	 & 	$-0.359$	 & 	$-0.370$	 & 	$0.069$	 & 	$0.318$	 & 	$0.114$ \\ \cline{1-8}
d5	 & 	$0.00025$	 & 	$0.000$	 & 	$-0.492$	 & 	$0.000$	 & 	$0.355$	 & 	$0.001$	 & 	$-0.278$ \\ \cline{1-8}
d6	 & 	$0.00025$	 & 	$-0.298$	 & 	$-0.359$	 & 	$0.370$	 & 	$0.069$	 & 	$-0.317$	 & 	$0.114$ \\ \cline{1-8}
d7	 & 	$0.00026$	 & 	$-0.299$	 & 	$-0.359$	 & 	$0.370$	 & 	$0.068$	 & 	$-0.317$	 & 	$0.115$ \\ \cline{1-8}
d8	 & 	$0.00027$	 & 	$0.000$	 & 	$-0.491$	 & 	$0.000$	 & 	$0.353$	 & 	$0.000$	 & 	$-0.275$ \\ \cline{1-8}
d9	 & 	$0.00025$	 & 	$0.000$	 & 	$-0.492$	 & 	$0.000$	 & 	$0.355$	 & 	$0.000$	 & 	$-0.278$ \\ \cline{1-8}
d10	 & 	$0.00025$	 & 	$0.298$	 & 	$-0.359$	 & 	$-0.370$	 & 	$0.069$	 & 	$0.317$	 & 	$0.115$ \\ \cline{1-8}
d11	 & 	$0.00025$	 & 	$0.298$	 & 	$-0.359$	 & 	$-0.370$	 & 	$0.069$	 & 	$0.318$	 & 	$0.115$ \\ \cline{1-8}
d12	 & 	$0.00025$	 & 	$0.000$	 & 	$-0.492$	 & 	$0.000$	 & 	$0.354$	 & 	$0.000$	 & 	$-0.277$ \\ \cline{1-8}
d13	 & 	$0.00025$	 & 	$-0.298$	 & 	$-0.359$	 & 	$0.370$	 & 	$0.069$	 & 	$-0.318$	 & 	$0.114$ \\ \cline{1-8}
d14	 & 	$0.00026$	 & 	$-0.299$	 & 	$-0.359$	 & 	$0.370$	 & 	$0.069$	 & 	$-0.317$	 & 	$0.115$ \\ \cline{1-8}
d15	 & 	$0.00028$	 & 	$-0.804$	 & 	$0.473$	 & 	$-0.107$	 & 	$-0.193$	 & 	$0.353$	 & 	$-0.354$ \\ \cline{1-8}
d16	 & 	$0.00011$	 & 	$-0.308$	 & 	$-0.353$	 & 	$0.381$	 & 	$0.055$	 & 	$-0.323$	 & 	$0.135$ \\ \cline{1-8}
d17	 & 	$0.00028$	 & 	$0.307$	 & 	$-0.350$	 & 	$-0.376$	 & 	$0.053$	 & 	$0.313$	 & 	$0.130$ \\ \cline{1-8}
d18	 & 	$0.00028$	 & 	$0.804$	 & 	$0.474$	 & 	$0.107$	 & 	$-0.192$	 & 	$-0.353$	 & 	$-0.354$ \\ \cline{1-8}
d19	 & 	$0.00028$	 & 	$0.804$	 & 	$0.473$	 & 	$0.107$	 & 	$-0.193$	 & 	$-0.353$	 & 	$-0.354$ \\ \cline{1-8}
d20	 & 	$0.00028$	 & 	$0.307$	 & 	$-0.350$	 & 	$-0.376$	 & 	$0.053$	 & 	$0.314$	 & 	$0.130$ \\ \cline{1-8}
d21	 & 	$0.00028$	 & 	$-0.307$	 & 	$-0.350$	 & 	$0.376$	 & 	$0.054$	 & 	$-0.314$	 & 	$0.129$ \\ \cline{1-8}
d22	 & 	$0.00028$	 & 	$-0.804$	 & 	$0.474$	 & 	$-0.107$	 & 	$-0.192$	 & 	$0.353$	 & 	$-0.354$ \\ \cline{1-8}
	\hline \hline
	$\epsilon^{\rm pk} $	&	$0.00562$ & \multicolumn{6}{|c|}{} \\ \cline{1-8}
	\hline
	\end{tabular}
	\caption{Numerical values of $\bar{P}_{d,p}$ for $1 \le p \le 6$ for $E=8$ MeV.}
	\label{jikkou8MeV}
\end{table}
\end{landscape}
\begin{landscape}
\begin{table}[htbp]
	\centering
		\small
	\begin{tabular}{|c || c | c | c | c | c | c | c | }
	\hline
	$E_\gamma=9\,[{\rm MeV}]$ & $\epsilon^{{\rm pk},w}_{d}(E_\gamma)$ & $\bar{P}_{d,1}$ & $\bar{P}_{d,2}$ & $\bar{P}_{d,3}$ & $\bar{P}_{d,4}$ & $\bar{P}_{d,5}$ & $\bar{P}_{d,6}$  \\ \hline
d1	 & 	$0.00023$	 & 	$0.000$	 & 	$-0.491$	 & 	$0.000$	 & 	$0.354$	 & 	$0.000$	 & 	$-0.276$ \\ \cline{1-8}
d2	 & 	$0.00021$	 & 	$0.000$	 & 	$-0.492$	 & 	$0.000$	 & 	$0.355$	 & 	$0.000$	 & 	$-0.279$ \\ \cline{1-8}
d3	 & 	$0.00022$	 & 	$0.298$	 & 	$-0.359$	 & 	$-0.370$	 & 	$0.069$	 & 	$0.318$	 & 	$0.115$ \\ \cline{1-8}
d4	 & 	$0.00021$	 & 	$0.298$	 & 	$-0.359$	 & 	$-0.370$	 & 	$0.069$	 & 	$0.318$	 & 	$0.115$ \\ \cline{1-8}
d5	 & 	$0.00021$	 & 	$0.000$	 & 	$-0.492$	 & 	$0.000$	 & 	$0.355$	 & 	$0.000$	 & 	$-0.279$ \\ \cline{1-8}
d6	 & 	$0.00021$	 & 	$-0.299$	 & 	$-0.359$	 & 	$0.370$	 & 	$0.069$	 & 	$-0.318$	 & 	$0.115$ \\ \cline{1-8}
d7	 & 	$0.00022$	 & 	$-0.298$	 & 	$-0.359$	 & 	$0.370$	 & 	$0.069$	 & 	$-0.318$	 & 	$0.115$ \\ \cline{1-8}
d8	 & 	$0.00023$	 & 	$0.000$	 & 	$-0.491$	 & 	$0.000$	 & 	$0.353$	 & 	$0.000$	 & 	$-0.275$ \\ \cline{1-8}
d9	 & 	$0.00021$	 & 	$0.000$	 & 	$-0.492$	 & 	$0.000$	 & 	$0.355$	 & 	$0.000$	 & 	$-0.278$ \\ \cline{1-8}
d10	 & 	$0.00021$	 & 	$0.298$	 & 	$-0.359$	 & 	$-0.370$	 & 	$0.069$	 & 	$0.318$	 & 	$0.114$ \\ \cline{1-8}
d11	 & 	$0.00021$	 & 	$0.298$	 & 	$-0.359$	 & 	$-0.370$	 & 	$0.069$	 & 	$0.318$	 & 	$0.115$ \\ \cline{1-8}
d12	 & 	$0.00021$	 & 	$0.000$	 & 	$-0.492$	 & 	$0.000$	 & 	$0.355$	 & 	$0.000$	 & 	$-0.278$ \\ \cline{1-8}
d13	 & 	$0.00021$	 & 	$-0.298$	 & 	$-0.359$	 & 	$0.370$	 & 	$0.069$	 & 	$-0.318$	 & 	$0.114$ \\ \cline{1-8}
d14	 & 	$0.00022$	 & 	$-0.299$	 & 	$-0.359$	 & 	$0.370$	 & 	$0.069$	 & 	$-0.318$	 & 	$0.115$ \\ \cline{1-8}
d15	 & 	$0.00024$	 & 	$-0.804$	 & 	$0.474$	 & 	$-0.107$	 & 	$-0.193$	 & 	$0.354$	 & 	$-0.355$ \\ \cline{1-8}
d16	 & 	$0.00009$	 & 	$-0.308$	 & 	$-0.353$	 & 	$0.381$	 & 	$0.055$	 & 	$-0.324$	 & 	$0.135$ \\ \cline{1-8}
d17	 & 	$0.00024$	 & 	$0.307$	 & 	$-0.351$	 & 	$-0.376$	 & 	$0.054$	 & 	$0.314$	 & 	$0.130$ \\ \cline{1-8}
d18	 & 	$0.00024$	 & 	$0.804$	 & 	$0.474$	 & 	$0.107$	 & 	$-0.193$	 & 	$-0.354$	 & 	$-0.355$ \\ \cline{1-8}
d19	 & 	$0.00024$	 & 	$0.804$	 & 	$0.473$	 & 	$0.107$	 & 	$-0.193$	 & 	$-0.354$	 & 	$-0.355$ \\ \cline{1-8}
d20	 & 	$0.00024$	 & 	$0.307$	 & 	$-0.350$	 & 	$-0.376$	 & 	$0.054$	 & 	$0.314$	 & 	$0.130$ \\ \cline{1-8}
d21	 & 	$0.00024$	 & 	$-0.307$	 & 	$-0.350$	 & 	$0.376$	 & 	$0.053$	 & 	$-0.314$	 & 	$0.130$ \\ \cline{1-8}
d22	 & 	$0.00024$	 & 	$-0.804$	 & 	$0.473$	 & 	$-0.106$	 & 	$-0.194$	 & 	$0.354$	 & 	$-0.355$ \\ \cline{1-8}
	\hline \hline
	$\epsilon^{\rm pk} $	&	$0.00478$ & \multicolumn{6}{|c|}{} \\ \cline{1-8}
	\hline
	\end{tabular}
	\caption{Numerical values of $\bar{P}_{d,p}$ for $1 \le p \le 6$ for $E=9$ MeV.}
	\label{jikkou9MeV}
\end{table}
\end{landscape}
\begin{landscape}
\begin{table}[htbp]
	\centering
		\small
	\begin{tabular}{|c || c | c | c | c | c | c | c | }
	\hline
	$E_\gamma=10\,[{\rm MeV}]$ & $\epsilon^{{\rm pk},w}_{d}(E_\gamma)$ & $\bar{P}_{d,1}$ & $\bar{P}_{d,2}$ & $\bar{P}_{d,3}$ & $\bar{P}_{d,4}$ & $\bar{P}_{d,5}$ & $\bar{P}_{d,6}$  \\ \hline
d1	 & 	$0.00020$	 & 	$0.000$	 & 	$-0.491$	 & 	$0.000$	 & 	$0.354$	 & 	$0.000$	 & 	$-0.276$ \\ \cline{1-8}
d2	 & 	$0.00018$	 & 	$0.000$	 & 	$-0.492$	 & 	$0.000$	 & 	$0.356$	 & 	$0.000$	 & 	$-0.279$ \\ \cline{1-8}
d3	 & 	$0.00018$	 & 	$0.299$	 & 	$-0.359$	 & 	$-0.370$	 & 	$0.069$	 & 	$0.318$	 & 	$0.115$ \\ \cline{1-8}
d4	 & 	$0.00018$	 & 	$0.298$	 & 	$-0.359$	 & 	$-0.370$	 & 	$0.070$	 & 	$0.319$	 & 	$0.115$ \\ \cline{1-8}
d5	 & 	$0.00018$	 & 	$0.000$	 & 	$-0.492$	 & 	$0.000$	 & 	$0.356$	 & 	$0.000$	 & 	$-0.279$ \\ \cline{1-8}
d6	 & 	$0.00018$	 & 	$-0.298$	 & 	$-0.359$	 & 	$0.370$	 & 	$0.069$	 & 	$-0.319$	 & 	$0.115$ \\ \cline{1-8}
d7	 & 	$0.00019$	 & 	$-0.298$	 & 	$-0.359$	 & 	$0.370$	 & 	$0.069$	 & 	$-0.319$	 & 	$0.115$ \\ \cline{1-8}
d8	 & 	$0.00020$	 & 	$0.000$	 & 	$-0.491$	 & 	$0.000$	 & 	$0.354$	 & 	$0.000$	 & 	$-0.276$ \\ \cline{1-8}
d9	 & 	$0.00018$	 & 	$0.000$	 & 	$-0.492$	 & 	$0.000$	 & 	$0.355$	 & 	$0.000$	 & 	$-0.279$ \\ \cline{1-8}
d10	 & 	$0.00017$	 & 	$0.298$	 & 	$-0.360$	 & 	$-0.370$	 & 	$0.070$	 & 	$0.319$	 & 	$0.114$ \\ \cline{1-8}
d11	 & 	$0.00018$	 & 	$0.298$	 & 	$-0.359$	 & 	$-0.370$	 & 	$0.069$	 & 	$0.319$	 & 	$0.115$ \\ \cline{1-8}
d12	 & 	$0.00018$	 & 	$0.000$	 & 	$-0.492$	 & 	$0.000$	 & 	$0.355$	 & 	$0.000$	 & 	$-0.279$ \\ \cline{1-8}
d13	 & 	$0.00018$	 & 	$-0.298$	 & 	$-0.360$	 & 	$0.370$	 & 	$0.070$	 & 	$-0.319$	 & 	$0.114$ \\ \cline{1-8}
d14	 & 	$0.00018$	 & 	$-0.298$	 & 	$-0.359$	 & 	$0.370$	 & 	$0.069$	 & 	$-0.319$	 & 	$0.115$ \\ \cline{1-8}
d15	 & 	$0.00021$	 & 	$-0.804$	 & 	$0.474$	 & 	$-0.107$	 & 	$-0.193$	 & 	$0.354$	 & 	$-0.356$ \\ \cline{1-8}
d16	 & 	$0.00008$	 & 	$-0.308$	 & 	$-0.353$	 & 	$0.381$	 & 	$0.055$	 & 	$-0.324$	 & 	$0.136$ \\ \cline{1-8}
d17	 & 	$0.00020$	 & 	$0.307$	 & 	$-0.350$	 & 	$-0.376$	 & 	$0.053$	 & 	$0.314$	 & 	$0.130$ \\ \cline{1-8}
d18	 & 	$0.00020$	 & 	$0.804$	 & 	$0.473$	 & 	$0.107$	 & 	$-0.193$	 & 	$-0.354$	 & 	$-0.356$ \\ \cline{1-8}
d19	 & 	$0.00020$	 & 	$0.804$	 & 	$0.474$	 & 	$0.107$	 & 	$-0.193$	 & 	$-0.354$	 & 	$-0.356$ \\ \cline{1-8}
d20	 & 	$0.00020$	 & 	$0.307$	 & 	$-0.351$	 & 	$-0.376$	 & 	$0.054$	 & 	$0.315$	 & 	$0.130$ \\ \cline{1-8}
d21	 & 	$0.00020$	 & 	$-0.307$	 & 	$-0.351$	 & 	$0.376$	 & 	$0.054$	 & 	$-0.315$	 & 	$0.130$ \\ \cline{1-8}
d22	 & 	$0.00020$	 & 	$-0.804$	 & 	$0.474$	 & 	$-0.107$	 & 	$-0.193$	 & 	$0.354$	 & 	$-0.356$ \\ \cline{1-8}
	\hline \hline
	$\epsilon^{\rm pk} $	&	$0.00406$ & \multicolumn{6}{|c|}{} \\ \cline{1-8}
	\hline
	\end{tabular}
	\caption{Numerical values of $\bar{P}_{d,p}$ for $1 \le p \le 6$ for $E=10$ MeV.}
	\label{jikkou10MeV}
\end{table}
\end{landscape}
\begin{landscape}
\begin{table}[htbp]
	\centering
		\small
	\begin{tabular}{|c || c | c | c | c | c | c | c | }
	\hline
	$E_\gamma=11\,[{\rm MeV}]$ & $\epsilon^{{\rm pk},w}_{d}(E_\gamma)$ & $\bar{P}_{d,1}$ & $\bar{P}_{d,2}$ & $\bar{P}_{d,3}$ & $\bar{P}_{d,4}$ & $\bar{P}_{d,5}$ & $\bar{P}_{d,6}$  \\ \hline
d1	 & 	$0.00017$	 & 	$0.000$	 & 	$-0.492$	 & 	$0.000$	 & 	$0.354$	 & 	$0.000$	 & 	$-0.277$ \\ \cline{1-8}
d2	 & 	$0.00016$	 & 	$0.000$	 & 	$-0.492$	 & 	$0.000$	 & 	$0.356$	 & 	$0.000$	 & 	$-0.280$ \\ \cline{1-8}
d3	 & 	$0.00016$	 & 	$0.298$	 & 	$-0.359$	 & 	$-0.370$	 & 	$0.070$	 & 	$0.319$	 & 	$0.115$ \\ \cline{1-8}
d4	 & 	$0.00015$	 & 	$0.299$	 & 	$-0.359$	 & 	$-0.371$	 & 	$0.069$	 & 	$0.319$	 & 	$0.116$ \\ \cline{1-8}
d5	 & 	$0.00015$	 & 	$0.000$	 & 	$-0.492$	 & 	$0.000$	 & 	$0.356$	 & 	$0.000$	 & 	$-0.280$ \\ \cline{1-8}
d6	 & 	$0.00015$	 & 	$-0.298$	 & 	$-0.360$	 & 	$0.370$	 & 	$0.070$	 & 	$-0.319$	 & 	$0.115$ \\ \cline{1-8}
d7	 & 	$0.00016$	 & 	$-0.298$	 & 	$-0.360$	 & 	$0.370$	 & 	$0.070$	 & 	$-0.319$	 & 	$0.115$ \\ \cline{1-8}
d8	 & 	$0.00017$	 & 	$0.000$	 & 	$-0.492$	 & 	$0.000$	 & 	$0.354$	 & 	$0.000$	 & 	$-0.277$ \\ \cline{1-8}
d9	 & 	$0.00015$	 & 	$0.000$	 & 	$-0.492$	 & 	$0.000$	 & 	$0.356$	 & 	$0.000$	 & 	$-0.279$ \\ \cline{1-8}
d10	 & 	$0.00015$	 & 	$0.298$	 & 	$-0.360$	 & 	$-0.370$	 & 	$0.070$	 & 	$0.319$	 & 	$0.114$ \\ \cline{1-8}
d11	 & 	$0.00015$	 & 	$0.298$	 & 	$-0.360$	 & 	$-0.370$	 & 	$0.070$	 & 	$0.319$	 & 	$0.114$ \\ \cline{1-8}
d12	 & 	$0.00015$	 & 	$0.000$	 & 	$-0.492$	 & 	$0.000$	 & 	$0.356$	 & 	$0.000$	 & 	$-0.280$ \\ \cline{1-8}
d13	 & 	$0.00016$	 & 	$-0.298$	 & 	$-0.359$	 & 	$0.371$	 & 	$0.070$	 & 	$-0.319$	 & 	$0.115$ \\ \cline{1-8}
d14	 & 	$0.00016$	 & 	$-0.298$	 & 	$-0.359$	 & 	$0.371$	 & 	$0.069$	 & 	$-0.319$	 & 	$0.115$ \\ \cline{1-8}
d15	 & 	$0.00018$	 & 	$-0.804$	 & 	$0.474$	 & 	$-0.107$	 & 	$-0.194$	 & 	$0.355$	 & 	$-0.357$ \\ \cline{1-8}
d16	 & 	$0.00006$	 & 	$-0.308$	 & 	$-0.353$	 & 	$0.381$	 & 	$0.055$	 & 	$-0.324$	 & 	$0.136$ \\ \cline{1-8}
d17	 & 	$0.00017$	 & 	$0.307$	 & 	$-0.351$	 & 	$-0.376$	 & 	$0.054$	 & 	$0.315$	 & 	$0.130$ \\ \cline{1-8}
d18	 & 	$0.00017$	 & 	$0.805$	 & 	$0.474$	 & 	$0.108$	 & 	$-0.193$	 & 	$-0.354$	 & 	$-0.357$ \\ \cline{1-8}
d19	 & 	$0.00017$	 & 	$0.805$	 & 	$0.474$	 & 	$0.107$	 & 	$-0.193$	 & 	$-0.355$	 & 	$-0.357$ \\ \cline{1-8}
d20	 & 	$0.00017$	 & 	$0.307$	 & 	$-0.351$	 & 	$-0.377$	 & 	$0.054$	 & 	$0.315$	 & 	$0.130$ \\ \cline{1-8}
d21	 & 	$0.00017$	 & 	$-0.307$	 & 	$-0.351$	 & 	$0.377$	 & 	$0.054$	 & 	$-0.315$	 & 	$0.130$ \\ \cline{1-8}
d22	 & 	$0.00017$	 & 	$-0.804$	 & 	$0.473$	 & 	$-0.106$	 & 	$-0.194$	 & 	$0.355$	 & 	$-0.356$ \\ \cline{1-8}
	\hline \hline
	$\epsilon^{\rm pk} $	&	$0.00345$ & \multicolumn{6}{|c|}{} \\ \cline{1-8}
	\hline
	\end{tabular}
	\caption{Numerical values of $\bar{P}_{d,p}$ for $1 \le p \le 6$ for $E=11$ MeV.}
	\label{jikkou11MeV}
\end{table}
\end{landscape}
\begin{landscape}
\begin{table}[htbp]
	\centering
		\small
	\begin{tabular}{|c || c | c | c | c | c | c | c | }
	\hline
	$E_\gamma=12\,[{\rm MeV}]$ & $\epsilon^{{\rm pk},w}_{d}(E_\gamma)$ & $\bar{P}_{d,1}$ & $\bar{P}_{d,2}$ & $\bar{P}_{d,3}$ & $\bar{P}_{d,4}$ & $\bar{P}_{d,5}$ & $\bar{P}_{d,6}$  \\ \hline
d1	 & 	$0.00015$	 & 	$-0.001$	 & 	$-0.492$	 & 	$0.001$	 & 	$0.355$	 & 	$-0.001$	 & 	$-0.278$ \\ \cline{1-8}
d2	 & 	$0.00013$	 & 	$0.000$	 & 	$-0.492$	 & 	$0.000$	 & 	$0.356$	 & 	$0.000$	 & 	$-0.280$ \\ \cline{1-8}
d3	 & 	$0.00014$	 & 	$0.298$	 & 	$-0.360$	 & 	$-0.371$	 & 	$0.070$	 & 	$0.320$	 & 	$0.115$ \\ \cline{1-8}
d4	 & 	$0.00013$	 & 	$0.298$	 & 	$-0.360$	 & 	$-0.371$	 & 	$0.070$	 & 	$0.320$	 & 	$0.115$ \\ \cline{1-8}
d5	 & 	$0.00013$	 & 	$0.000$	 & 	$-0.492$	 & 	$0.000$	 & 	$0.356$	 & 	$0.000$	 & 	$-0.280$ \\ \cline{1-8}
d6	 & 	$0.00013$	 & 	$-0.298$	 & 	$-0.360$	 & 	$0.370$	 & 	$0.070$	 & 	$-0.320$	 & 	$0.114$ \\ \cline{1-8}
d7	 & 	$0.00013$	 & 	$-0.298$	 & 	$-0.360$	 & 	$0.371$	 & 	$0.070$	 & 	$-0.320$	 & 	$0.115$ \\ \cline{1-8}
d8	 & 	$0.00014$	 & 	$0.000$	 & 	$-0.492$	 & 	$0.000$	 & 	$0.354$	 & 	$0.000$	 & 	$-0.277$ \\ \cline{1-8}
d9	 & 	$0.00013$	 & 	$0.000$	 & 	$-0.492$	 & 	$0.000$	 & 	$0.356$	 & 	$0.001$	 & 	$-0.280$ \\ \cline{1-8}
d10	 & 	$0.00013$	 & 	$0.298$	 & 	$-0.360$	 & 	$-0.371$	 & 	$0.070$	 & 	$0.319$	 & 	$0.115$ \\ \cline{1-8}
d11	 & 	$0.00013$	 & 	$0.298$	 & 	$-0.360$	 & 	$-0.371$	 & 	$0.070$	 & 	$0.320$	 & 	$0.115$ \\ \cline{1-8}
d12	 & 	$0.00013$	 & 	$0.000$	 & 	$-0.492$	 & 	$0.000$	 & 	$0.356$	 & 	$0.000$	 & 	$-0.280$ \\ \cline{1-8}
d13	 & 	$0.00013$	 & 	$-0.298$	 & 	$-0.360$	 & 	$0.370$	 & 	$0.070$	 & 	$-0.319$	 & 	$0.114$ \\ \cline{1-8}
d14	 & 	$0.00013$	 & 	$-0.298$	 & 	$-0.359$	 & 	$0.371$	 & 	$0.069$	 & 	$-0.320$	 & 	$0.116$ \\ \cline{1-8}
d15	 & 	$0.00015$	 & 	$-0.805$	 & 	$0.474$	 & 	$-0.107$	 & 	$-0.193$	 & 	$0.355$	 & 	$-0.357$ \\ \cline{1-8}
d16	 & 	$0.00005$	 & 	$-0.308$	 & 	$-0.353$	 & 	$0.381$	 & 	$0.055$	 & 	$-0.325$	 & 	$0.136$ \\ \cline{1-8}
d17	 & 	$0.00015$	 & 	$0.307$	 & 	$-0.351$	 & 	$-0.377$	 & 	$0.054$	 & 	$0.315$	 & 	$0.130$ \\ \cline{1-8}
d18	 & 	$0.00015$	 & 	$0.805$	 & 	$0.474$	 & 	$0.108$	 & 	$-0.193$	 & 	$-0.355$	 & 	$-0.358$ \\ \cline{1-8}
d19	 & 	$0.00015$	 & 	$0.805$	 & 	$0.474$	 & 	$0.107$	 & 	$-0.193$	 & 	$-0.355$	 & 	$-0.357$ \\ \cline{1-8}
d20	 & 	$0.00015$	 & 	$0.307$	 & 	$-0.351$	 & 	$-0.377$	 & 	$0.054$	 & 	$0.315$	 & 	$0.131$ \\ \cline{1-8}
d21	 & 	$0.00015$	 & 	$-0.307$	 & 	$-0.351$	 & 	$0.377$	 & 	$0.054$	 & 	$-0.315$	 & 	$0.131$ \\ \cline{1-8}
d22	 & 	$0.00015$	 & 	$-0.805$	 & 	$0.474$	 & 	$-0.107$	 & 	$-0.193$	 & 	$0.355$	 & 	$-0.357$ \\ \cline{1-8}
	\hline \hline
	$\epsilon^{\rm pk} $	&	$0.00295$ & \multicolumn{6}{|c|}{} \\ \cline{1-8}
	\hline
	\end{tabular}
	\caption{Numerical values of $\bar{P}_{d,p}$ for $1 \le p \le 6$ for $E=12$ MeV.}
	\label{jikkou12MeV}
\end{table}
\end{landscape}

\section{Summary}
In this paper, a Monte Carlo simulation with GEANT4 toolkit is presented, to reproduce the measurements using radioactive sources and a melamine target at the ANNRI. 
Using a simulation, we calculated the energy dependence of $\epsilon^{{\rm pk}}$ and $\bar{P}_{d,p}$. 
This novel method can be applied in the study of the partial neutron widths of p-wave resonances in the entrance channel to the compound states, which is required in the study of the discrete symmetry breaking in compound states.
The germanium detector assembly installed at the ANNRI was characterized for the measurement of the angular distribution of individual $\gamma$-rays.








\section*{\label{sec:level3}Acknowledgement}
We thanks to the staff of the ANNRI for the maintenance of the germanium detectors, and the MLF and the J-PARC for operating the accelerators.
The measurement was performed under the User Program (Proposal No. 2014S03 and 2015S12) of the MLF in the J-PARC.
This work was supported by MEXT KAKENHI Grant number JP19GS0210 and JSPS KAKENHI Grant Number JP17H02889.

\end{document}